\documentclass[12pt]{article}
\usepackage{graphicx}
\usepackage{lscape}
\usepackage{epsfig}
\usepackage{citesort}
\usepackage{amssymb}
\usepackage{amsmath}
\newcommand{\be}{\begin{equation}}
\newcommand{\ee}{\end{equation}}
\newcommand{\bea}{\begin{eqnarray}}
\newcommand{\eea}{\end{eqnarray}}

\def\R1{\varepsilon_1}
\def\E8{\varepsilon_8}

\def\s1{\hat s}

\newcommand{\bd}{\begin{displaymath}}
\newcommand{\ed}{\end{displaymath}}

\def\R1{\varepsilon_1}
\def\E8{\varepsilon_8}

\def\beq{\begin{equation}}
\def\eeq{\end{equation}}
\def\bea{\begin{eqnarray}}
\def\eea{\end{eqnarray}}
\def\beeq{\begin{eqnarray}}
\def\eeeq{\end{eqnarray}}

\def\ba{\begin{array}}
\def\ea{\end{array}}

\def\xis0{{\Xi^{*0}}}

\def\g5{\gamma_5}

\begin{document}
\title{
         {\Large
                 {\bf Positive and negative parity hyperons in nuclear medium
                 }
         }
      }
\author{\vspace{1cm}\\
{\small  K. Azizi$^1$ \thanks {e-mail: kazizi@dogus.edu.tr}\,\,, N. Er$^{1,2}$\thanks {e-mail: nuray@ibu.edu.tr}, H.
Sundu$^3$\thanks {e-mail: hayriye.sundu@kocaeli.edu.tr}} \\
{\small $^1$ Department of Physics, Do\u gu\c s University,
Ac{\i}badem-Kad{\i}k\"oy, 34722 \.{I}stanbul, Turkey}\\
{\small $^2$ Department of Physics, Abant \.{I}zzet Baysal University,
G\"olk\"oy Kamp\"us\"u, 14980 Bolu, Turkey}\\
{\small $^3$ Department of Physics, Kocaeli University,
 41380 \.{I}zmit, Turkey}\\
}
\date{}

\begin{titlepage}
\maketitle
\thispagestyle{empty}
\begin{abstract}
The effects of nuclear medium on the residue, mass and self energy of the positive and negative parity $\Sigma$, $\Lambda$ and $\Xi$ hyperons are investigated 
using the QCD sum rule method. In the calculations, the general interpolating currents of hyperons with an arbitrary mixing parameter 
are used. We compare the results obtained in medium with those of the vacuum and calculate the shifts in the corresponding parameters. 
It is found that the shifts on the residues in nuclear matter are over all positive  for both the positive and negative parity  hyperons, except for the positive parity  $\Sigma$ hyperon
 that the shift is negative. The shifts on the masses of 
these baryons are obtained to be  negative. 
The shifts on the residues and masses of negative parity states are large compared to those of positive parities. The maximum shift belongs to the residue of the negative parity  $\Lambda$ hyperon.
The vector self-energies gained by the positive parity baryons are large compared to the negative parities' vector self-energies. The maximum value of the vector self-energy belongs to the positive 
parity $\Sigma$ hyperon. The numerical values are compared with the existing predictions in the literature. 
\end{abstract}
~~~PACS number(s): 14.20.Jn,  21.65.-f, 11.55.Hx
\end{titlepage}

%%%%%%%%%%%%%%%%%%%%%%%%%%%%%%%%%%%%%%%%%%%%%%%%%%%%%%%%%

\section{Introduction}

The study of in medium properties of hadrons constitutes one of the main research directions in QCD.
It helps us gain valuable knowledge on the structure of dense astrophysical objects like neutron stars,
QCD phase diagram, as well as perturbative and non-perturbative natures of QCD. The investigation of in-medium hadronic
parameters can also help us in analyzing the results of heavy  ion collision experiments. The effects of nuclear medium on 
nucleon parameters have been widely investigated in the literature \cite{Drukarev88,Drukarev90,Hatsuda91,Adami91,kanur2014}. 
But, there are a few number
of works dedicated to the study of these effects on the properties of strange members of the octet baryons. 
It is well known that the nucleons are affected by the nuclear medium considerably \cite{kanur2014}. 
As  $\Sigma, \Lambda$ and $\Xi$  hyperons have also  u and d quarks content their inside, it is expected 
that their parameters are also affected by the medium. The investigation of hyperons in nuclear medium and 
the comparison of the results with the nucleons can helps us to determine the order of $SU(3)_f$ violation in medium. 
The predictions of the scalar and vector self-energies of these particles can also provide valuable 
information on the scalar and vector in-medium couplings of these particles.  In \cite{Jin94}, the authors 
study the self-energies of the $\Lambda$ hyperon in nuclear medium using the finite-density QCD sum rules.
Their calculations show that the vector and scalar self-energies of the $\Lambda$ hyperon are
substantially  smaller than the corresponding nucleon self-energies.
In \cite{Jin95}, the authors apply the same method to investigate the self-energies of the $\Sigma$ hyperon 
propagating in nuclear matter. They find that the Lorentz vector self-energy of  $\Sigma$ is similar to that of the nucleon.
They show that the magnitude of Lorentz scalar self-energy of the $\Sigma$ hyperon is close to the corresponding value of the nucleon; 
although it is sensitive to the strangeness content of the nucleon and to the density dependence of certain four-quark condensate. In \cite{Savage96}, 
the authors consider mass shifts for the baryon octet using a model independent approach to baryon-baryon interactions based on the chiral perturbation theory.
In \cite{Miyatsu2009}, the authors have applied the chiral quark-meson coupling (CQMC) model by including the effects of  gluon and pion exchanges  
to study the properties of hyperons in a nuclear matter.
In \cite{Beane2012}, S. R. Beane et al. use $n\Sigma^{-}$ scattering phase shifts to quantify the energy shift of  the  $\Sigma^{-}$ hyperon in dense neutron matter.
 
In the present work, we extend the above mentioned studies by including the negative parity hyperons. In particular, we use the QCD sum 
rule approach to calculate the shifts on the residues, masses and the self energies of   $\Sigma, \Lambda$ and $\Xi$ hyperons due to the nuclear medium for both positive and negative parities. We use the most general form of the interpolating currents of these baryons and try to find 
reliable region for the general parameter entering the interpolating fields. We compare the results obtained with existing predictions in the literature for the positive parity hyperons. 
 
The article is organized as follow.  In section 2, we obtain QCD sum rules for the residues, masses and self energies of 
the $\Sigma, \Lambda$ and $\Xi$ hyperons in the nuclear matter. Section 3 is devoted to the numerical analyses of the sum rules 
and the  comparison of the results with the existing predictions as well as with the vacuum results. Section 4 contains our concluding remarks.

%%%%%%%%%%%%%%%%%%%%%%%%%%%%%%%%%%%%%%%%%%%%%%%%%%%%%%%%%

\section{In-medium QCD sum rules for the residues, masses and self energies of hyperons }

To obtain the QCD sum rules for  the residues , masses and self energies of $\Sigma, \Lambda, \Xi$  hyperons  in the presence of nuclear matter, the starting point is to consider the following two-point correlation function:
\begin{equation}\label{correilk}
\Pi(p)=i\int{d^4 xe^{ip\cdot x}\langle\psi_0|T[J_H
(x)\bar{J}_H(0)]|\psi_0\rangle},
\end{equation}
where $p$ is the four momentum of the hyperon, $|\psi_0\rangle$ is the nuclear matter ground
state and the subindex  $H$  is used for the $\Sigma, \Lambda, \Xi$ hyperons. 
Here $J_H$ is the related interpolating current coupling to both the positive and negative parities. The  general form of $J_H$ for the corresponding baryons are taken as \cite{Chung,Dosch} 
\begin{eqnarray}\label{currents}
J_{\Sigma}(x)&=&-\frac{1}{\sqrt{2}}\epsilon_{abc} \sum_{i=1}^{2}\Big[\Big(u^{T,a}(x)CA_{1}^{i}s^{b}(x)\Big)A_{2}^{i}d^{c}(x)
-\Big(d^{T,c}(x)CA_{1}^{i}s^{b}(x)\Big)A_{2}^{i}u^{a}(x)\Big], \nonumber \\
J_{\Lambda}(x)&=&\frac{1}{\sqrt{6}}\epsilon_{abc}\sum_{i=1}^{2}\Big[2\Big(u^{T,a}(x)CA_{1}^{i}d^{b}(x)\Big)A_{2}^{i}s^{c}(x)
+\Big(u^{T,a}(x)CA_{1}^{i}s^{b}(x)\Big)A_{2}^{i}d^{c}(x)\nonumber \\
&&+\Big(d^{T,c}(x)CA_{1}^{i}s^{b}(x)\Big)A_{2}^{i}u^{a}(x)\Big], \nonumber \\
J_{\Xi}(x)&=&-\epsilon_{abc}\sum_{i=1}^{2}\Big(s^{T,a}(x)CA_{1}^{i}u^{b}(x)\Big)A_{2}^{i}s^{c}(x),
\end{eqnarray}
where $a, b, c$ are color indices, $C$ is the charge conjugation operator and
$A_{1}^{1}=I$, $A_{1}^{2}=A_{2}^{1}=\gamma_5$, 
$A_{2}^{2}=\beta$. As previously said, the parameter $\beta$ is an arbitrary auxiliary parameter,
and $\beta=-1$ corresponds to the Ioffe current \cite{Drukarev2013,Thomas,Leinweber,Stein}. 
Considering the Lorentz covariance and parity invariance the correlation function can be decomposed as 
\begin{equation}
\Pi(p)=\Pi_{p}\!\not\!{p}+\Pi_{u}\!\not\!{u}+ \Pi_{S} I+\Pi^{\prime}(p_{\mu}u_{\nu}-p_{\nu}u_{\mu})\sigma^{\mu\nu},
\end{equation}
where the $\Pi_{i}$'s and $\Pi^{\prime}$ are the functions of the invariants $p^2$ and $p\cdot u$ with $u$ being the four vector velocity of the nuclear medium. In the calculations, we will use the structures $\!\not\!{p}, \!\not\!{u}$ and unit matrix $I$ to construct the sum rules for the quantities under consideration. 
According to the method used, we calculate the aforesaid correlation function in two hadronic and OPE sides. Matching these two sides through a dispersion relation leads to the sum rules for the considered parameters.

%%%%%%%%%%%%%%%%%%%%%%%%%%%%%%%%%%%%%%%%%%%%%%%%%%%%%%%%%%

\subsection{Hadronic representation}

The correlation function can be calculated  by inserting  complete sets of hyperon states with both 
parities and with the same quantum numbers as the interpolating currents.  After performing integral over four-x, we get
\begin{eqnarray}\label{corre}
\Pi^{Had}=&-&\frac{\langle\psi_0|J_{H^{+}}(x)|H^{+}(p^*,s)\rangle\langle H^{+}(p^*,s)|\bar{J}_{H^{+}}(0)|\psi_0\rangle}{p^{*2}-m_{H^{+}}^{*2}} \nonumber \\
&-&\frac{\langle\psi_0|J_{H^{-}}(x)|H^{-}(p^*,s)\rangle\langle H^{-}(p^*,s)|\bar{J}_{H^{-}}(0)|\psi_0\rangle}{p^{*2}-m_{H^{-}}^{*2}} + ... ,
\end{eqnarray}
where $|H^{+}(p^*,s)\rangle$ and $|H^{-}(p^*,s)\rangle$ are the hyperon states with positive and negative parity, respectively and  the dots represents the contributions of the higher states and the continuum. Here $p^*$ is the four momentum  and $m_{H}^{*}$ is the mass of the hyperon in medium. The matrix elements appearing in Eq. (\ref{corre}) can be parametrized as
\begin{eqnarray}\label{intcur}
\langle\psi_0|J_{H^{+}}(x)|H^+(p^*,s)\rangle&=&\lambda_{H^{+}}^{*}u_{H^+}(p^*,s) , \nonumber \\
\langle\psi_0|J_{H^{-}}(x)|H^-(p^*,s)\rangle&=&\lambda_{H^{-}}^{*}\gamma_5 u_{H^-}(p^*,s),
\end{eqnarray}
where $\lambda_{H^{+}}^{*}$ and  $\lambda_{H^{-}}^{*}$ are the modified residues or the coupling strengths of the positive and negative parity hyperons, respectively; $u_{H^+}(p^*,s)$ and $u_{H^-}(p^*,s)$ are their Dirac spinors. Using  Eq. (\ref{intcur}) in Eq. (\ref{corre}) 
and summing over the spins of positive and negative parity hyperons, we get 
\begin{equation}
\Pi^{Had}=-\frac{\lambda_{H^{+}}^{*2}(\!\not\!{p^*}+m_{H^+}^{*})}{p^{*2}-m_{H^+}^{*2}}-\frac{\lambda_{H^{-}}^{*2}(\!\not\!{p^*}-m_{H^-}^{*})}{p^{*2}-m_{H^-}^{*2}}+... ,
\end{equation}
which can be written as 
\begin{eqnarray}\label{corre2}
\Pi^{Had}&=&-\frac{\lambda_{H^{+}}^{*2}}{\!\not\!p^{*}-m_{H^+}^{*}}-\frac{\lambda_{H^{-}}^{*2}}{\!\not\!p^{*}+m_{H^-}^{*}}+... \nonumber \\
&=&-\frac{\lambda_{H^{+}}^{*2}}{
(p^{\mu}_{H^+}-\Sigma_{\nu H^+}^{\mu})\gamma_\mu-(m_{H^+}+\Sigma^{S}_{H^+})} -\frac{\lambda_{H^{-}}^{*2}}{
(p^{\mu}_{H^-}-\Sigma_{\nu H^-}^{\mu})\gamma_\mu+(m_{H^-}+\Sigma^{S}_{H^-})} +... . \nonumber \\
\end{eqnarray}
where $\Sigma_{\nu H^{\pm}}^{\mu}$ and $\Sigma^{S}_{H^{\pm}}$ are the vector and the scalar self-energies of the positive and negative 
parity hyperons in nuclear matter, respectively \cite{TDC2}. 
In general, we can write 
\begin{equation}\label{sigma1}
\Sigma_{\nu H^{\pm}}^{\mu}=\Sigma_{\nu H^{\pm}} u^\mu+\Sigma'_{\nu H^{\pm}}p_{H^{\pm}}^\mu,
\end{equation}
where $\Sigma_{\nu H^{\pm}}$ and $\Sigma'_{\nu H^{\pm}}$ are constants and $ u^\mu$ is the four velocity of the nuclear medium. 
Here, we neglect $\Sigma'_{\nu H^{\pm}}$ due to its small contribution  (see also \cite{TDC2}).
Apart from the vacuum QCD calculations, the four-velocity of the nuclear matter is
new concept that causes extra structures to the correlation function. We shall work in the
rest frame of the hyperon with $u^\mu=(1,0)$. Substitution of Eq. (\ref{sigma1}) into Eq. (\ref{corre2}), 
the hadronic side of the correlation function  can be written in terms of three different structure as 
\begin{equation}
\Pi^{Had}=\Pi^{Had}_p(p^2,p_0)\!\not\!{p}+\Pi^{Had}_u(p^2,p_0)\!\not\!{u}
+\Pi^{Had}_S(p^2,p_0)I+ ...,
\end{equation}
where $p_0$ is the energy of the quasi-particle and
\begin{eqnarray}
\Pi^{Had}_p(p^2,p_0)&=&-\lambda_{H^{+}}^{*2}\frac{1}{p^2-\mu_{H^+}^2}-\lambda_{H^{-}}^{*2}\frac{1}{p^2-\mu_{H^-}^2},\nonumber \\
\Pi^{Had}_u(p^2,p_0)&=&+\lambda_{H^{+}}^{*2}\frac{\Sigma_{H^+\nu}}{p^2-\mu_{H^+}^2}+\lambda_{H^{-}}^{*2}\frac{\Sigma_{H^-\nu}}{p^2-\mu_{H^-}^2},
\nonumber \\
\Pi^{Had}_S(p^2,p_0)&=&-\lambda_{H^{+}}^{*2}\frac{m_{H^+}^*}{p^2-\mu_{H^+}^2} +\lambda_{H^{-}}^{*2}\frac{m_{H^-}^*}{p^2-\mu_{H^-}^2}.
\end{eqnarray}
Here $m_{H^{\pm}}^*=m_{H^{\pm}}+\Sigma^{S}_{H^{\pm}}$ and
$\mu_{H^{\pm}}^2=m_{H^{\pm}}^{*2}-\Sigma_{\nu H^{\pm}}^2+2p^{H^{\pm}}_{0}\Sigma_{\nu H^{\pm}}$. After a Wick rotation and applying the Borel transformation with respect to $p^2$, we get 
\begin{eqnarray}\label{3eqn6unknown}
\hat{B}\Pi^{Had}_p(p^2,p_0)&=&\lambda_{H^{+}}^{*2}e^{-\mu_{H^+}^2/M^2}+\lambda_{H^{-}}^{*2}e^{-\mu_{H^-}^2/M^2}, \nonumber \\
\hat{B}\Pi^{Had}_u(p^2,p_0)&=&-\lambda_{H^{+}}^{*2}\Sigma_{\nu H^+} e^{-\mu_{H^+}^2/M^2}-\lambda_{H^{-}}^{*2}\Sigma_{\nu H^-} e^{-\mu_{H^-}^2/M^2},
\nonumber \\
\hat{B}\Pi^{Had}_S(p^2,p_0)&=&\lambda_{H^{+}}^{*2} m_{H^+}^* e^{-\mu_{H^+}^2/M^2}-\lambda_{H^{-}}^{*2} m_{H^-}^* e^{-\mu_{H^-}^2/M^2},
\end{eqnarray}
where $M^2$ is the Borel mass parameter.

%%%%%%%%%%%%%%%%%%%%%%%%%%%%%%%%%%%%%%%%%%%%%%%%%%%%%%%%%%

\subsection{OPE side}

The OPE side of the correlation function can be calculated  in deep Euclidean region. It can also be 
written in terms of the considered structures  as 
\begin{equation}
\Pi^{OPE}(p)=\Pi_{p}^{OPE}\!\not\! {p}+\Pi_{u}^{OPE}\!\not\!
{u}+\Pi_{S}^{OPE}I+... .
\end{equation}
Our main task in the following is to calculate the $\Pi_i^{OPE}$ functions, where $i= \!\not\! {p}, \!\not\! {u}$ and $I$. Using the explicit forms of the interpolating currents in the correlation function  in Eq. (\ref{correilk}) and contracting out
all quark pairs via Wick's theorem, we find
\begin{eqnarray}\label{sigma}
\Pi^{OPE}_{\Sigma}(p) &=&
\frac{i}{2}\epsilon_{abc}\epsilon_{a'b'c'}\int d^4 x e^{ipx}\Big\langle \psi_0\Big
|\Big\{\Big(\gamma_{5}S^{ca'}_{d}(x)S'^{bb'}_{s}(x)S^{ac'}_{u}(x)\gamma_{5}
\nonumber \\
&+&\gamma_{5}S^{cb'}_{u}(x)S'^{aa'}_{s}(x)S^{bc'}_{d}(x)\gamma_{5}
+\gamma_{5}S^{cc'}_{u}(x)\gamma_{5}Tr\Big[S^{ab'}_{s}(x)S'^{ba'}_{d}(x)\Big] \nonumber \\
&+&\gamma_{5}S^{cc'}_{d}(x)\gamma_{5}Tr\Big[S^{ab'}_{u}(x)S'^{ba'}_{s}(x)\Big] \Big)
+\beta\Big(\gamma_{5}S^{ca'}_{d}(x)\gamma_{5}S'^{bb'}_{s}(x)S^{ac'}_{u}
(x)\nonumber \\
&+&\gamma_{5}S^{cb'}_{u}(x)\gamma_{5}S'^{aa'}_{s}(x)S^{bc'}_{d}(x)
+S^{ca'}_{d}(x)S'^{bb'}_{s}(x)\gamma_{5}S^{ac'}_{u}(x)\gamma_{5}
\nonumber \\
&+&S^{cb'}_{u}(x)S'^{aa'}_{s}(x)\gamma_{5}S^{bc'}_{d}(x)\gamma_{5}
+\gamma_{5}S^{cc'}_{u}(x)Tr\Big[S^{ab'}_{s}(x)\gamma_{5}S'^{ba'}_{d}(x)\Big] \nonumber \\
&+&S^{cc'}_{u}(x)\gamma_{5}Tr\Big[S^{ab'}_{s}(x)S'^{ba'}_{d}(x)\gamma_{5}\Big]
+\gamma_{5}S^{cc'}_{d}(x)Tr\Big[S^{ab'}_{u}(x)\gamma_{5}S'^{ba'}_{s}(x)\Big] \nonumber \\
&+&S^{cc'}_{d}(x)\gamma_{5}Tr\Big[S^{ab'}_{u}(x)S'^{ba'}_{s}(x)\gamma_{5}\Big]
\Big)
+\beta^2\Big(S^{ca'}_{d}(x)\gamma_{5}S'^{bb'}_{s}(x)\gamma_{5}S^{ac'}_{u}
(x)\nonumber \\
&+&S^{cb'}_{u}(x)\gamma_{5}S'^{aa'}_{s}(x)\gamma_{5}S^{bc'}_{d}(x)
+S^{cc'}_{u}(x)Tr\Big[S^{ba'}_{d}(x)\gamma_{5}S'^{ab'}_{s}(x)\gamma_{5}\Big] \nonumber \\
&+&S^{cc'}_{d}(x)Tr\Big[S^{ba'}_{s}(x)\gamma_{5}S'^{ab'}_{u}(x)\gamma_{5}\Big]
\Big)
\Big\}\Big| \psi_0\Big\rangle,
\end{eqnarray}

\begin{eqnarray}\label{lamda}
\Pi^{OPE}_{\Lambda}(p) &=&
-\frac{i}{6}\epsilon_{abc}\epsilon_{a'b'c'}\int d^4 x e^{ipx}\Big\langle \psi_0\Big
|\Big\{\Big(\gamma_{5}S^{ca'}_{d}(x)S'^{bb'}_{s}(x)S^{ac'}_{u}(x)\gamma_{5}
\nonumber \\
&+&2\gamma_{5}S^{ca'}_{d}(x)S'^{ab'}_{u}(x)S^{bc'}_{s}(x)\gamma_{5}
+2\gamma_{5}S^{ca'}_{s}(x)S'^{ab'}_{u}(x)S^{bc'}_{d}(x)\gamma_{5}\nonumber \\
&+&2\gamma_{5}S^{cb'}_{s}(x)S'^{ba'}_{d}(x)S^{ac'}_{u}(x)\gamma_{5}
+2\gamma_{5}S^{cb'}_{u}(x)S'^{ba'}_{d}(x)S^{ac'}_{s}(x)\gamma_{5}\nonumber \\
&+&\gamma_{5}S^{cb'}_{u}(x)S'^{aa'}_{s}(x)S^{bc'}_{d}(x)\gamma_{5}
-\gamma_{5}S^{cc'}_{u}(x)\gamma_{5}Tr\Big[S^{ab'}_{s}(x)S'^{ba'}_{d}(x)\Big] \nonumber \\
&-&4\gamma_{5}S^{cc'}_{s}(x)\gamma_{5}Tr\Big[S^{ab'}_{u}(x)S'^{ba'}_{d}(x)\Big] 
-\gamma_{5}S^{cc'}_{d}(x)\gamma_{5}Tr\Big[S^{ab'}_{u}(x)S'^{ba'}_{s}(x)\Big]\Big) \nonumber \\
&+&\beta\Big(\gamma_{5}S^{ca'}_{d}(x)\gamma_{5}S'^{bb'}_{s}(x)S^{ac'}_{u}(x)
+2\gamma_{5}S^{ca'}_{d}(x)\gamma_{5}S'^{ab'}_{u}(x)S^{bc'}_{s}(x)\nonumber \\
&+&2\gamma_{5}S^{ca'}_{s}(x)\gamma_{5}S'^{ab'}_{u}(x)S^{bc'}_{d}(x)
+2\gamma_{5}S^{cb'}_{s}(x)\gamma_{5}S'^{ba'}_{d}(x)S^{ac'}_{u}(x) \nonumber \\
&+&2\gamma_{5}S^{cb'}_{u}(x)\gamma_{5}S'^{ba'}_{d}(x)S^{ac'}_{s}(x)
+\gamma_{5}S^{cb'}_{u}(x)\gamma_{5}S'^{aa'}_{s}(x)S^{bc'}_{d}(x) \nonumber \\
&+&S^{ca'}_{d}(x)S'^{bb'}_{s}(x)\gamma_{5}S^{ac'}_{u}(x)\gamma_{5}
+2S^{ca'}_{d}(x)S'^{ab'}_{u}(x)\gamma_{5}S^{bc'}_{s}(x)\gamma_{5} \nonumber \\
&+&2S^{ca'}_{s}(x)S'^{ab'}_{u}(x)\gamma_{5}S^{bc'}_{d}(x)\gamma_{5} 
+2S^{cb'}_{s}(x)S'^{ba'}_{d}(x)\gamma_{5}S^{ac'}_{u}(x)\gamma_{5} \nonumber \\
&+&2S^{cb'}_{u}(x)S'^{ba'}_{d}(x)\gamma_{5}S^{ac'}_{s}(x)\gamma_{5}
+S^{cb'}_{u}(x)S'^{aa'}_{s}(x)\gamma_{5}S^{bc'}_{d}(x)\gamma_{5} \nonumber \\
&-&\gamma_{5}S^{cc'}_{u}(x)Tr\Big[S^{ab'}_{s}(x)\gamma_{5}S'^{ba'}_{d}(x)\Big]
-S^{cc'}_{u}(x)\gamma_{5}Tr\Big[S^{ab'}_{s}(x)S'^{ba'}_{d}(x)\gamma_{5}\Big]  \nonumber \\
&-&4\gamma_{5}S^{cc'}_{s}(x)Tr\Big[S^{ab'}_{u}(x)\gamma_{5}S'^{ba'}_{d}(x)\Big] 
-\gamma_{5}S^{cc'}_{d}(x)Tr\Big[S^{ab'}_{u}(x)\gamma_{5}S'^{ba'}_{s}(x)\Big]  \nonumber \\
&-&4S^{cc'}_{s}(x)\gamma_{5}Tr\Big[S^{ab'}_{u}(x)S'^{ba'}_{d}(x)\gamma_{5}\Big] 
-S^{cc'}_{d}(x)\gamma_{5}Tr\Big[S^{ab'}_{u}(x)S'^{ba'}_{s}(x)\gamma_{5}\Big]\Big)  \nonumber 
\end{eqnarray}
\begin{eqnarray}
&+&\beta^2\Big(S^{ca'}_{d}(x)\gamma_{5}S'^{bb'}_{s}(x)\gamma_{5}S^{ac'}_{u}(x)
+2S^{ca'}_{d}(x)\gamma_{5}S'^{ab'}_{u}(x)\gamma_{5}S^{bc'}_{s}(x) \nonumber \\
&+&2S^{ca'}_{s}(x)\gamma_{5}S'^{ab'}_{u}(x)\gamma_{5}S^{bc'}_{d}(x)\Big)
+2S^{cb'}_{s}(x)\gamma_{5}S'^{ba'}_{d}(x)\gamma_{5}S^{ac'}_{u}(x) \nonumber \\
&+& 2S^{cb'}_{u}(x)\gamma_{5}S'^{ba'}_{d}(x)\gamma_{5}S^{ac'}_{s}(x)
+S^{cb'}_{u}(x)\gamma_{5}S'^{aa'}_{s}(x)\gamma_{5}S^{bc'}_{d}(x) \nonumber \\
&-&S^{cc'}_{u}(x)Tr\Big[S^{ba'}_{d}(x)\gamma_{5}S'^{ab'}_{s}(x)\gamma_{5}\Big]
-4S^{cc'}_{s}(x)Tr\Big[S^{ba'}_{d}(x)\gamma_{5}S'^{ab'}_{u}(x)\gamma_{5}\Big] \nonumber \\
&-&S^{cc'}_{d}(x)Tr\Big[S^{ba'}_{s}(x)\gamma_{5}S'^{ab'}_{u}(x)\gamma_{5}\Big]
\Big\}\Big| \psi_0\Big\rangle,
\end{eqnarray}

\begin{eqnarray}\label{Xi}
\Pi^{OPE}_{\Xi}(p) &=&
i\epsilon_{abc}\epsilon_{a'b'c'}\int d^4 x e^{ipx}\Big\langle \psi_0\Big
|\Big\{\Big(-\gamma_{5}S^{cb'}_{s}(x)S'^{ba'}_{u}(x)S^{ac'}_{s}(x)\gamma_{5}
\nonumber \\
&+&\gamma_{5}S^{cc'}_{s}(x)\gamma_{5}Tr\Big[S^{ab'}_{s}(x)S'^{ba'}_{u}(x)\Big]\Big)
-\beta\Big(\gamma_{5}S^{cb'}_{s}(x)\gamma_{5}S'^{ba'}_{u}(x)S^{ac'}_{s}(x)\nonumber \\
&+&S^{cb'}_{s}(x)S'^{ba'}_{u}(x)\gamma_{5}S^{ac'}_{s}(x)\gamma_{5}
-\gamma_{5}S^{cc'}_{s}(x)Tr\Big[S^{ab'}_{s}(x)\gamma_{5}S'^{ba'}_{u}(x)\Big] \nonumber \\
&-&S^{cc'}_{s}(x)\gamma_{5}Tr\Big[S^{ab'}_{s}(x)S'^{ba'}_{u}(x)\gamma_{5}\Big]\Big)
+\beta^2\Big(-S^{cb'}_{s}(x)\gamma_{5}S'^{ba'}_{u}(x)\gamma_{5}S^{ac'}_{s}(x)\nonumber \\
&+&S^{cc'}_{s}(x)Tr\Big[S^{ba'}_{u}(x)\gamma_{5}S'^{ab'}_{s}(x)\gamma_{5}\Big]\Big)
\Big\}\Big| \psi_0\Big\rangle,
\end{eqnarray}
where $S'=CS^TC$, $S_{u,d, s}$ are light quarks propagators and $Tr[...]$ denotes the trace of gamma matrices.  
In coordinate-space, the light quark
propagator at the nuclear medium has the following form in the fixed-point gauge \cite{Cohen}:
\begin{eqnarray}\label{propagator}
 S_{q}^{ab}(x)&\equiv& \langle\psi_0|T[q^a
(x)\bar{q}^b(0)]|\psi_0\rangle_{\rho_N} \nonumber \\
&=&
\frac{i}{2\pi^2}\delta^{ab}\frac{1}{(x^2)^2}\not\!x
-\frac{m_q }{ 4\pi^2} \delta^ { ab } \frac { 1}{x^2} \nonumber \\
&+&
\chi^a_q(x)\bar{\chi}^b_q(0)-\frac{ig_s}{32\pi^2}F_{\mu\nu}^A(0)t^{ab,A
}\frac{1}{x^2}[\not\!x\sigma^{\mu\nu}+\sigma^{\mu\nu}\not\!x]+...,
\end{eqnarray}
where $\rho_N$ is the nuclear matter density, $\chi^a_q$ and $\bar{\chi}^b_q$ are the Grassmann background quark fields and
$F_{\mu\nu}^A$ are classical background gluon fields (for details see  \cite{Cohen}).

The invariant $\Pi^{OPE}_{i}(p)$ functions in the Borel scheme for each hyperon can be written  in terms of the  perturbative part and non-perturbative parts up to dimension six as following
\begin{eqnarray} \label{OPE}
\widehat{\textbf{B}}\Pi^{OPE}_{i}(p)=\widehat{\textbf{B}}\Pi{_i}^{Pert}+\widehat{\textbf{B}}\Pi_{i,D3}^{Non-pert}+\widehat{\textbf{B}}\Pi_{i,D4}^{Non-pert}
+\widehat{\textbf{B}}\Pi_{i,D5}^{Non-pert}+\widehat{\textbf{B}}\Pi_{i,D6}^{Non-pert}.
\end{eqnarray}
The perturbative parts for all structures and the considered hyperons are found as 
\begin{eqnarray} 
\left.\begin{array}{l}\widehat{\textbf{B}}\Pi^{Pert}_{p}=-\frac{1}{2048\pi^4}[5+2\beta+5\beta^2]\int^{s_0}_{0}s^2e^{-\frac{s}{M^2}}ds \\ 
\widehat{\textbf{B}}\Pi^{Pert}_{u}= 0 \\
\widehat{\textbf{B}}\Pi^{Pert}_{S}=-\frac{1}{512\pi^4}(\beta -1)[3(\beta+1)(m_u+m_d)+(\beta -1)m_s]\int^{s_0}_{0}s^2e^{-\frac{s}{M^2}}ds
\end{array}
\right\} \textrm{for} ~ \Sigma, \nonumber \\
\end{eqnarray} 
\begin{eqnarray} 
\left.\begin{array}{l}\widehat{\textbf{B}}\Pi^{Pert}_{p}=-\frac{1}{2048\pi^4}[5+\beta(2+5\beta)]\int^{s_0}_{0}s^2e^{-\frac{s}{M^2}}ds\\
\widehat{\textbf{B}}\Pi^{Pert}_{u}= 0 \\
\widehat{\textbf{B}}\Pi^{Pert}_{S}=-\frac{1}{1536\pi^4}(\beta -1)[(5\beta+1)(m_u+m_d)+(11\beta +13)m_s]\int^{s_0}_{0}s^2e^{-\frac{s}{M^2}}ds
\end{array}
\right\} \textrm{for} ~ \Lambda, \nonumber \\
\end{eqnarray} 
\begin{eqnarray} 
\left.\begin{array}{l}\widehat{\textbf{B}}\Pi^{Pert}_{p}=-\frac{1}{2048\pi^4}[5+\beta(2+5\beta)]\int^{s_0}_{0}s^2e^{-\frac{s}{M^2}}ds \\
\widehat{\textbf{B}}\Pi^{Pert}_{u}= 0 \\
\widehat{\textbf{B}}\Pi^{Pert}_{S}= -\frac{1}{512\pi^4}(\beta -1)[(\beta-1)m_u+6(\beta+1)m_s]\int^{s_0}_{0}s^2e^{-\frac{s}{M^2}}ds\end{array}
\right\} \textrm{for} ~ \Xi.\nonumber \\
\end{eqnarray} 

The functions $\widehat{\textbf{B}}\Pi_{i,D3}^{Non-pert}, \widehat{\textbf{B}}\Pi_{i,D4}^{Non-pert}, \widehat{\textbf{B}}\Pi_{i,D5}^{Non-pert}, 
\widehat{\textbf{B}}\Pi_{i,D6}^{Non-pert}$ in non-perturbative parts have very lengthy expressions. So, we only present the function $\widehat{\textbf{B}}\Pi_{i,D3}^{Non-pert}$
for the structure $\!\not\!{p}$ and $\Sigma$ particle, as an example. It is obtained as 
\begin{eqnarray} \label{Dim3}
\widehat{\textbf{B}}\Pi_{\Sigma,D3}^{Non-pert}&=&\frac{\langle
\bar{d}g_s\sigma G d\rangle_{\rho_N}+\langle \bar{u}g_s\sigma G
u\rangle_{\rho_N}}{128 \pi^2
M^2}(1-\beta^2)m_s(3M^2-2p_0^2)+\frac{\langle d^{\dag} iD_0iD_0
d\rangle_{\rho_N}}{48\pi^2}\nonumber \\
&\times& p_0\Big[(3+2\beta+3\beta^2)M^2-2(1+\beta^2)p_0^2\Big]+\frac{\langle
d^{\dag}g_s \sigma Gd\rangle_{\rho_N}+\langle u^{\dag}g_s \sigma
Gu\rangle_{\rho_N}}{576\pi^2 M^2} \nonumber \\
&\times& p_0(1+\beta^2)(2p_0^2-3M^2) +\frac{\langle s^{\dag}g_s
\sigma Gs\rangle_{\rho_N}}{576\pi^2
M^2}p_0(3+2\beta+3\beta^2)(2p_0^2-3M^2) \nonumber \\
&+&\frac{\langle \bar{d}iD_0iD_0 d \rangle_{\rho_N}+\langle
\bar{u}iD_0iD_0 u \rangle_{\rho_N}}{8\pi^2
M^2}p_0^2m_s(1-\beta^2) \nonumber \\
&-& \frac{\langle s^{\dag}iD_0iD_0 s\rangle_{\rho_N}}{48 \pi^2
M^2} p_0\Big[M^2(1+\beta)^2+2p_0^2(3+2\beta+3\beta^2)\Big] \nonumber \\
&+& \frac{\langle u^{\dag}iD_0iD_0 u\rangle_{\rho_N}}{48 \pi^2
M^2} p_0\Big[M^2(3+2\beta+3\beta^2)-2p_0^2(1+\beta^2)\Big] \nonumber \\
&-&\frac{i\langle d^{\dag}iD_0 d\rangle_{\rho_N}+i\langle
u^{\dag}iD_0 u\rangle_{\rho_N}}{288\pi^2}\Big[8p_0^2(1+\beta^2)-
(11+6\beta+11\beta^2)\int^{s_0}_{0}e^{-\frac{s}{M^2}}ds\Big] \nonumber \\
&-&\frac{i\langle s^{\dag}iD_0
s\rangle_{\rho_N}}{288\pi^2}\Big[8p_0^2(3+2\beta+3\beta^2)-
(3-2\beta+3\beta^2)\int^{s_0}_{0}e^{-\frac{s}{M^2}}ds\Big] \nonumber \\
&+&3\frac{\langle \bar{d}d\rangle_{\rho_N}+\langle
\bar{u}u\rangle_{\rho_N}}{64\pi^2}m_s(-1+\beta^2)\int^{s_0}_{0}e^{-\frac{s}{M^2}}ds \nonumber \\
&+&\frac{\langle d^{\dag}d\rangle_{\rho_N}+\langle
u^{\dag}u\rangle_{\rho_N}}{96\pi^2}p_0(1+\beta^2)\int^{s_0}_{0}e^{-\frac{s}{M^2}}ds \nonumber \\
&+&\frac{\langle s^{\dag}s\rangle_{\rho_N}}{96\pi^2}p_0(3+2\beta+3\beta^2)\int^{s_0}_{0}e^{-\frac{s}{M^2}}ds,
\end{eqnarray}
where $s_0$ is the continuum threshold and we ignored to present the terms containing the up and down quark masses. Equating the hadronic and OPE sides of the correlation
functions, we get
\begin{eqnarray}\label{3eqn8unknownOPE}
\lambda_{H^{+}}^{*2}e^{-\mu_{H^+}^2/M^2}+\lambda_{H^{-}}^{*2}e^{-\mu_{H^-}^2/M^2} &=& \Pi^{OPE}_p, \nonumber \\
-\lambda_{H^{+}}^{*2}\Sigma_{\nu H^+} e^{-\mu_{H^+}^2/M^2}-\lambda_{H^{-}}^{*2}\Sigma_{\nu H^-} e^{-\mu_{H^-}^2/M^2} &=& \Pi^{OPE}_u,
\nonumber \\
\lambda_{H^{+}}^{*2} m_{H^+}^* e^{-\mu_{H^+}^2/M^2}-\lambda_{H^{-}}^{*2} m_{H^-}^* e^{-\mu_{H^-}^2/M^2} &=& \Pi^{OPE}_S.
\end{eqnarray}
To find the eight unknowns $\lambda_{H^{+}}^{*}, \lambda_{H^{-}}^{*}, m_{H^+}^{*}, m_{H^-}^{*}, \Sigma_{\nu H^+}, \Sigma_{\nu H^-}, \mu_{H^+}$ and $\mu_{H^-}$, 
we need five more equations
which are found by applying derivatives with respect to $(-1/M^2)$ to both sides of  Eq. (\ref{3eqn8unknownOPE}). As a result, we get
\begin{eqnarray}\label{3eqn8unknownOPEiki}
\lambda_{H^{+}}^{*2} \mu_{H^+}^2 e^{-\mu_{H^+}^2/M^2}+\lambda_{H^{-}}^{*2} \mu_{H^-}^2 e^{-\mu_{H^-}^2/M^2} &=& \frac{d \Pi^{OPE}_p}{d(-1/M^2)}, \nonumber \\
-\lambda_{H^{+}}^{*2}\Sigma_{\nu H^+} \mu_{H^+}^2 e^{-\mu_{H^+}^2/M^2}-\lambda_{H^{-}}^{*2}\Sigma_{\nu H^-} \mu_{H^-}^2 e^{-\mu_{H^-}^2/M^2} &=&\frac{d \Pi^{OPE}_u}{d(-1/M^2)},
\nonumber \\
\lambda_{H^{+}}^{*2} m_{H^+}^* \mu_{H^+}^2 e^{-\mu_{H^+}^2/M^2}-\lambda_{H^{-}}^{*2} m_{H^-}^* \mu_{H^-}^2 e^{-\mu_{H^-}^2/M^2} &=& \frac{d \Pi^{OPE}_S}{d(-1/M^2)}, \nonumber \\
\lambda_{H^{+}}^{*2} (\mu_{H^+}^2)^2 e^{-\mu_{H^+}^2/M^2}+\lambda_{H^{-}}^{*2} (\mu_{H^-}^2)^2 e^{-\mu_{H^-}^2/M^2}&=& \frac{d^2 \Pi^{OPE}_p}{d(-1/M^2)^2}, \nonumber \\
-\lambda_{H^{+}}^{*2}\Sigma_{\nu H^+} (\mu_{H^+}^2)^2 e^{-\mu_{H^+}^2/M^2}-\lambda_{H^{-}}^{*2}\Sigma_{\nu H^-} (\mu_{H^-}^2)^2 e^{-\mu_{H^-}^2/M^2}&=& \frac{d^2 \Pi^{OPE}_u}{d(-1/M^2)^2}.
\end{eqnarray}
By solving the above eight equations simultaneously, we get the following in-medium sum rules for the parameters under consideration 
\begin{eqnarray}
\lambda_{H^{+}}^{*2}&=&\Bigg[ \frac{-2 Q_{2}^{2} Q_{3} + 2 Q_{1} Q_{2} Q_{4} + Q_{1} Q_{3} Q_{7} - Q_{1}^{2} Q_{8} + Q_{1} \tilde{Q}}{2 \tilde{Q}}\Bigg]e^{\mu_{H^+}^2/M^2}, \nonumber \\
\lambda_{H^{-}}^{*2}&=& \Bigg[ \frac{-2 Q_{2}^{2} Q_{3} + 2 Q_{1} Q_{2} Q_{4} + Q_{1} Q_{3} Q_{7} - Q_{1}^{2} Q_{8} + Q_{1} \tilde{Q}}{2 \tilde{Q}}\Bigg]e^{\mu_{H^-}^2/M^2}, \nonumber \\
m_{H^{+}}^{*}&=&\Bigg[ \frac{2 Q_{2} Q_{3} Q_{6} - 2 Q_{1} Q_{4} Q_{6} - Q_{3} Q_{5} Q_{7} + Q_{1} Q_{5} Q_{8}-Q_{5} \tilde{Q}}{2 Q_{2}^2 Q_{3} - 2 Q_{1} Q_{2}Q_{4}-Q_{1} Q_{3} Q_{7} + Q_{1}^2 Q_{8} - Q_{1} \tilde{Q}}  \Bigg]\nonumber, \\
m_{H^{-}}^{*}&=&\Bigg[ \frac{-2 Q_{2} Q_{3} Q_{6} + 2 Q_{1} Q_{4} Q_{6} + Q_{3} Q_{5} Q_{7} - Q_{1} Q_{5} Q_{8}+Q_{5} \tilde{Q}}{2 Q_{2}^2 Q_{3} - 2 Q_{1} Q_{2}Q_{4}-Q_{1} Q_{3} Q_{7} + Q_{1}^2 Q_{8} - Q_{1} \tilde{Q}} \Bigg] \nonumber, \\
\Sigma_{\nu H^{\pm}}&=&\Bigg[ \frac{-2 Q_2 Q_4 + Q_3 Q_7 + Q_1 Q_8 + \tilde{Q}}{2 (Q_{2}^{2} - Q_1 Q_7)} \Bigg] \nonumber, \\
\mu_{H^{+}}^{*2}&=&\Bigg[ \frac{ Q_1 Q_8 -Q_3 Q_7  + \tilde{Q}}{2 (Q_1 Q_4-Q_2 Q_3)}\Bigg] \nonumber, \\
\mu_{H^{-}}^{*2}&=&\Bigg[ \frac{ Q_3 Q_7 - Q_1 Q_8 + \tilde{Q}}{2 (Q_2 Q_3 -  Q_1 Q_4)}\Bigg] \nonumber, \\
\end{eqnarray}
where 
\begin{eqnarray}\label{}
Q_1 = \Pi^{OPE}_p  ,  Q_2 = \frac{d \Pi^{OPE}_p}{d(-1/M^2)}  , \nonumber \\
Q_3 = \Pi^{OPE}_u  ,  Q_4 = \frac{d \Pi^{OPE}_u}{d(-1/M^2)}  , \nonumber \\
Q_5 = \Pi^{OPE}_S ,  Q_6 = \frac{d \Pi^{OPE}_S}{d(-1/M^2)}  , \nonumber \\
Q_7 = \frac{d^2 \Pi^{OPE}_p}{d(-1/M^2)^2} , Q_8 = \frac{d^2 \Pi^{OPE}_u}{d(-1/M^2)^2},
\end{eqnarray}
and $\tilde{Q}=\sqrt{(Q_{3} Q_{7} - Q_{1} Q_{8})^2 + 4 (Q_{2} Q_{3} - Q_{1} Q_{4}) ( Q_{2} Q_{8}-Q_{4} Q_{7} )}$. Here we shall remark that we choose the above roots for $\mu_{H^{\pm}}^{*2}$ among four roots obtained from the calculations.

%%%%%%%%%%%%%%%%%%%%%%%%%%%%%%%%%%%%%%%%%%%%%%%%%%%%%%%%%%

\section{Numerical results and discussion}
In this section, we numerically analyze the QCD sum rules for the residues, masses and self energies of $\Sigma, \Lambda$ and $\Xi$ hyperons for both parities in nuclear matter. 
For this aim, the numerical values of the quark masses as well as the in-medium quark-quark, 
quark-gluon, gluon-gluon condensates are necessary inputs. Their numerical values are listed in table 1. Besides, we need to find the working regions of three auxiliary parameters,  
namely continuum threshold $s_0$, Borel mass parameter $M^2$ and the arbitrary mixing parameter $\beta$. 
The standard criteria for finding the reliable working regions for these helping parameters is that, 
the physical quantities show good stability with respect to these parameters in those regions.  

The continuum threshold $s_0$ corresponds to the beginning of the continuum in the considered channels. 
$\sqrt{s_0}-m_H$ is the energy  that we need to excite the particle to its first excited state with the same quantum numbers. 
We take this value in the interval $[0.3-0.5]$ GeV. In this interval, the numerical results show weak dependency on the continuum threshold. 
To determine the working region for the Borel mass parameter $M^2$, we take into account two criteria: suppression of 
the contributions of the higher states and the continuum compared to the pole contribution and the exceeding of the perturbative 
part to the non-perturbative contributions.  More precisely, the upper bound on the Borel  parameter is found by demanding that 
\begin{eqnarray}
\label{nolabel}
\frac{ \int_{0}^{s_0}ds  \rho_{p,u,S}(s) e^{-s/M^2} }{
 \int_{0}^\infty ds \rho_{p,u,S}(s) e^{-s/M^2}} ~~>~~ 1/2, 
\end{eqnarray}
where $\rho_{p,u,S}(s)$ are the spectral densities corresponding to different structures in the channels under consideration. The lower bound on this parameter is obtained requiring that the perturbative  contribution is more than   the non-perturbative one
 and the term with highest dimension contributes less than 10\% to the whole OPE, i.e. the series of sum rules converge. 
Note that to find the Borel window we consider all the  sum rules in 
 Eqs. (\ref{3eqn8unknownOPE}) and  (\ref{3eqn8unknownOPEiki}) and choose the one with relatively worst OPE convergence. Our numerical calculations show that the last sum rule in
 Eq. (\ref{3eqn8unknownOPEiki}) has relatively worst OPE convergence but still satisfies the aforesaid criteria.
We consider this sum rule and the above mentioned conditions to determine the ``Borel Window`` at different channels.

In the case of the mixing parameter $\beta$, we  try 
to find a $\beta$ for each channel with a large residue/continuum ratio, showing  that the relative contribution of the ground state pole to the sum rules is largest. 
Considering these criteria and using all input parameters, in the following, we perform our numerical 
analyses for the hyperons under consideration. 
 
\begin{table}[ht!]
\centering
%\rowcolors{1}{lightgray}{white}
\begin{tabular}{ll}
\hline \hline
   Input parameters  &  Values    
           \\
\hline \hline
$p_0   $          &  $m_{H} $      \\
$ m_{u}   $ ; $ m_{d}   $ ;  $ m_{s}   $  &  $2.3_{-0.5}^{0.7}  $ $MeV$     ; $4.83_{-0.3}^{0.5}  $ $MeV$ ;   $95 \pm 5  $ $MeV$   \cite{PDG}      \\
$ m_{\Sigma^{+}} ; m_{\Sigma^{-}}  $  &  $1192.642 \pm 0.024  $ $MeV$ ; $ \approx 1620  $ $MeV$ \cite{PDG}  \\
$ m_{\Lambda^{+}} ; m_{\Lambda^{-}}  $  &  $1115.683 \pm 0.006  $ $MeV$ ; $1405.13_{-1.0}^{1.3}  $ $MeV$ \cite{PDG} \\
$ m_{\Xi^{+}} ; m_{\Xi^{-}}  $  &  $1314.86 \pm 0.2  $ $MeV$\cite{PDG} ; $1.8$ $GeV$ \cite{Yoshihiko}   \\
$ \rho_{N}     $          &  $(0.11)^3  $ $GeV^3$    \cite{Cohen,XJ1,Cohen45}     \\
$ \langle q^{\dag}q\rangle_{\rho_N}    $ ; $ \langle s^{\dag}s\rangle_{\rho_N}    $         &  $\frac{3}{2}\rho_{N}$   ; 0    \cite{Cohen,XJ1,Cohen45,Wang}   \\
$ \langle\bar{q}q\rangle_{0} $  ; $ \langle\bar{s}s\rangle_{0} $  &  $ (-0.241)^3    $ $GeV^3$ ; 0.8 $ \langle\bar{q}q\rangle_{0} $   \cite{Belyaev}        \\
$ m_{q}      $          &  $0.5(m_{u}+m_{d})$           \cite{Cohen,XJ1,Cohen45}      \\
$ \sigma_{N} $  ;   $ \sigma_{N_0} $      &  $0.045 ~  $GeV$ $ ; $0.035 ~  $GeV$ $  \cite{Cohen,XJ1,Cohen45}  \\
$y$ & $0.04\pm0.02$ \cite{Thomas1}; $0.066\pm0.011\pm0.002$ \cite{Dinter}\\
$  \langle\bar{q}q\rangle_{\rho_N}  $ ;  $  \langle\bar{s}s\rangle_{\rho_N}  $  &  $ \langle\bar{q}q\rangle_{0}+\frac{\sigma_{N}}{2m_{q}}\rho_{N}$ 
; $ \langle\bar{s}s\rangle_{0}+y\frac{\sigma_{N}}{2m_{q}}\rho_{N}$    \cite{Cohen,XJ1,Cohen45,Nielsen,Wang}              \\
$  \langle q^{\dag}g_{s}\sigma Gq\rangle_{\rho_N}  $ ;  $  \langle s^{\dag}g_{s}\sigma Gs\rangle_{\rho_N} $ &  $ -0.33~GeV^2 \rho_{N}$ ; $ -y 0.33~GeV^2 \rho_{N}$ \cite{Cohen,XJ1,Cohen45,Nielsen,Wang}\\
$  \langle q^{\dag}iD_{0}q\rangle_{\rho_N}  $    ; $  \langle s^{\dag}iD_{0}s\rangle_{\rho_N}  $      &  $0.18 ~GeV \rho_{N}$  ;$ \frac{m_s \langle\bar{s}s\rangle_{\rho_N}}{4}
+0.02 ~GeV \rho_N$            \cite{Cohen,XJ1,Cohen45,Nielsen,Wang}    \\
$  \langle\bar{q}iD_{0}q\rangle_{\rho_N}  $    ; $  \langle\bar{s}iD_{0}s\rangle_{\rho_N}  $      &  $\frac{3}{2} m_q \rho_{N}\simeq0 $ ; 0 \cite{Cohen,XJ1,Cohen45,Nielsen,Wang}\\
$  m_{0}^{2}  $          &  $ 0.8~GeV^2$    \cite{Belyaev}               \\
$   \langle\bar{q}g_{s}\sigma Gq\rangle_{0} $ ; $   \langle\bar{s}g_{s}\sigma Gs\rangle_{0} $  &  $m_{0}^{2}\langle\bar{q}q\rangle_{0} $ ; $m_{0}^{2}\langle\bar{s}s\rangle_{0} $ \\
$  \langle\bar{q}g_{s}\sigma Gq\rangle_{\rho_N}  $  ;$  \langle\bar{s}g_{s}\sigma Gs\rangle_{\rho_N}  $ &  $\langle\bar{q}g_{s}\sigma Gq\rangle_{0}+3~GeV^2\rho_{N} $ ;
$\langle\bar{s}g_{s}\sigma Gs\rangle_{0}+3 y ~GeV^2\rho_{N} $\cite{Cohen,XJ1,Cohen45,Nielsen,Wang} \\
$ \langle  \bar{q}iD_{0}iD_{0}q\rangle_{\rho_N} $  ;$\langle \bar{s}iD_{0}iD_{0}s\rangle_{\rho_N} $  &  $ 0.3~GeV^2\rho_{N}-\frac{1}{8}\langle\bar{q}g_{s}\sigma Gq\rangle_{\rho_N}$ ; \\
& $ 0.3 y ~GeV^2\rho_{N}-\frac{1}{8}\langle\bar{s}g_{s}\sigma Gs\rangle_{\rho_N}$  \cite{Cohen,XJ1,Cohen45,Nielsen,Wang}\\
$  \langle q^{\dag}iD_{0}iD_{0}q\rangle_{\rho_N}  $ ;$  \langle s^{\dag}iD_{0}iD_{0}s\rangle_{\rho_N}$ & $0.031~GeV^2\rho_{N}-\frac{1}{12}\langle q^{\dag}g_{s}\sigma Gq\rangle_{\rho_N} $; \\
& $0.031 y~GeV^2\rho_{N}-\frac{1}{12}\langle s^{\dag}g_{s}\sigma Gs\rangle_{\rho_N} $ \cite{Cohen,XJ1,Cohen45,Nielsen,Wang}\\
$\langle \frac{\alpha_s}{\pi} G^{2}\rangle_{0}$ & $(0.33\pm0.04)^4~GeV^4$ \cite{Belyaev}\\
$\langle \frac{\alpha_s}{\pi} G^{2}\rangle_{\rho_N}$ & $\langle \frac{\alpha_s}{\pi} G^{2}\rangle_{0}-0.65~GeV \rho_N$ \cite{Cohen,XJ1,Cohen45}\\
 \hline \hline
\end{tabular}
\caption{Numerical values for input parameters. We use the average value $y=0.05$ to perform the numerical analyses.}
\end{table}

%%%%%%%%%%%%%%%%%%%%%%%%%% 

\subsection{$\Sigma$ hyperon}
The aforementioned criteria for determination of the working region for the Borel mass parameter 
lead to the region $1.3$ GeV$^2\leqslant M^2\leqslant 1.9$ GeV$^2$ for the $\Sigma$ hyperon. From the previously said considerations for the fixing of the 
mixing parameter $\beta$, we  find  that for  $ cos\theta=-0.76$, where $\theta=tan^{-1}(\beta)$, the relative contribution of the ground state pole residue  is largest  for the  $\Sigma$ hyperon.
%  Taking the average 
% value of $M^2$, we plot the dependence of residues of the positive and the negative parity $\Sigma$ hyperon with 
% respect to $cos\theta ~ (\beta=tan\theta)$ in the interval $[-1,1]$ at different fixed value of $s_0$ in figure 1. 
% From this figure, we read the regions $-0.8 \leqslant cos\theta \leqslant -0.3$ and $0.3 \leqslant cos\theta \leqslant 0.8$ 
% for the positive parity; $-0.9 \leqslant cos\theta \leqslant -0.4$ and $0.4 \leqslant cos\theta \leqslant 0.9$ for the negative 
% parity $\Sigma$ hyperon as reliable working regions. 
Using these values, we plot the dependence of the residues, masses and 
vector self-energies of the positive and negative parity $\Sigma$ particle on the Borel mass parameter in figures 1-3. These figures demonstrate good stabilities with respect 
to $M^2$ and show weak dependence on the continuum threshold $s_0$ at their working 
regions. Obtained from the figures 1-3, the average values of the residues, masses and vector self-energies of the positive and 
negative parity $\Sigma$ hyperons are depicted in table 2. Taking $\rho_N \rightarrow 0$ in the OPE side, we can also calculate the numerical values of the residues and masses in vacuum. 
The ratios of the residues and masses in nuclear matter to those of the vacuum together with the ratios of the self-energies to the 
vacuum masses are listed in table 3. In this table, we also present the predictions existing in the literature for  the positive parity baryon. 
A quick glance in this table shows that, there are considerable shifts in the parameters under consideration in nuclear matter compared to 
their vacuum values for both parities in $\Sigma$ channel. The shift on the residue of positive parity is negative but it is positive for the negative parity state. The shifts are negative for the masses
 of both parities. 
The obtained result for the ratio of the masses for the positive parity state in the present work is close to those of \cite{Jin95,Cohen} within the errors.  
For the $\Sigma_{\nu\Sigma^{+}}/m_{\Sigma^{+}}$, our result shows  a small difference with those of \cite{Jin95,Cohen}. This can 
be attributed to the different interpolating currents used in these works and the fact that, we take into accounts both the positive and 
negative parity baryons. When we compare the results of parameters of the positive and negative parity cases, we see that the absolute value of the shift in the 
residue of the negative parity is considerably more than that of the positive parity $\Sigma$. Meanwhile, the absolute value of the shift on the mass for the negative parity is 
also higher than that of the positive parity. The ratio   
$\Sigma_{\nu\Sigma^{+}}/m_{\Sigma^{+}}$ is roughly $1.8$ times more than the ratio  $\Sigma_{\nu\Sigma^{-}}/m_{\Sigma^{-}}$. 
Using the relation $m^{*}_{\Sigma}=m_{\Sigma}+\Sigma^{S}_{\Sigma}$, we find the scalar self energy of the positive and the 
negative parity $\Sigma$ to be $-0.324$ GeV  and $-0.748$ GeV, respectively. Our results for remaining parameters may 
be checked by other phenomenological approaches as well as the future experiments. 
\begin{table}[ht!]
\centering
%\rowcolors{1}{lightgray}{white}
\begin{tabular}{|c|c|c|c|c|c|}
\hline \hline
 $\lambda^{*}_{\Sigma^{+}}$ [GeV$^3]$ & $\lambda^{*}_{\Sigma^{-}}$  [GeV$^3]$ & $m^{*}_{\Sigma^{+}}$  [GeV] & $m^{*}_{\Sigma^{-}}$ 
 [GeV] & $\Sigma_{\nu \Sigma^{+}}$ [GeV] & $\Sigma_{\nu \Sigma^{-}}$ [GeV]  \\
 \hline
$ 0.013\pm 0.003$ & $ 0.027 \pm 0.006 $ & $  0.924\pm 0.330 $ & $ 0.779\pm 0.287 $ & $ 0.350\pm 0.035 $ & $ 0.228\pm 0.023 $ \\ 
 \hline \hline
\end{tabular}
\caption{The numerical values of residues, masses and self energies of $\Sigma$ hyperon with positive and negative parities.}
\end{table}
\begin{table}[ht!]
\centering
%\rowcolors{1}{lightgray}{white}
\begin{tabular}{|c|c|c|c|c|c|c|}
\hline \hline
 & $\lambda^{*}_{\Sigma^{+}}/\lambda_{\Sigma^{+}}$ & $\lambda^{*}_{\Sigma^{-}}/\lambda_{\Sigma^{-}}$ & $m^{*}_{\Sigma^{+}}/m_{\Sigma^{+}}$ 
 & $m^{*}_{\Sigma^{-}}/m_{\Sigma^{-}}$ & $\Sigma_{\nu \Sigma^{+}}/m_{\Sigma^{+}}$ &  $\Sigma_{\nu\Sigma^{-}}/m_{\Sigma^{-}}$ \\
 \hline
Present work & $0.94 \pm 0.22 $ & $1.47\pm 0.27 $&$ 0.74\pm 0.14 $ & $0.51\pm 0.09$&$ 0.28\pm 0.06$ &$ 0.15\pm 0.03$ \\ 
 \hline
\cite{Jin95,Cohen} &  &  &  0.78-0.85 & & 0.18-0.19  & \\
\hline \hline
\end{tabular}
\caption{ The ration of parameters for $\Sigma$ hyperon.}
\end{table}

At the end of this part, we would like to compare our results for the positive parity $\Sigma$ baryon with those of 
nucleon \cite{kanur2014}. In the case of mass,  the shift is negative for both the $\Sigma$ baryon and nucleon, 
although the value of shift in the case of $\Sigma$ is $5\%$ less than the nucleon. 
As far as, the residue of the positive parity of the $\Sigma$ hyperon is concerned, the shift is $-6\%$ while it is $-10\%$ for the nucleon \cite{kanur2014}.

%%%%%%%%%%%%% Residuee Nuc mat %%%%%%%%%%%%%

%\begin{figure}[h!]
%\label{fig1}
%\centering
%\begin{tabular}{ccc}
%\epsfig{file=SigResiduePozitiveParityNMx.eps,width=0.45\linewidth,clip=} &
%\epsfig{file=SigResidueNegativeParityNMx.eps,width=0.45\linewidth,clip=}
%\end{tabular}
%\caption{The residue of the positive parity $\Sigma$ hyperon versus $cos\theta$  in nuclear matter (left panel). The same for negative parity $\Sigma$ hyperon (right panel).}
%\end{figure}

%%%%%%%%%%%%% Residuee Nuc mat %%%%%%%%%%%%%

\begin{figure}[h!]
\label{fig1}
\centering
\begin{tabular}{ccc}
\epsfig{file=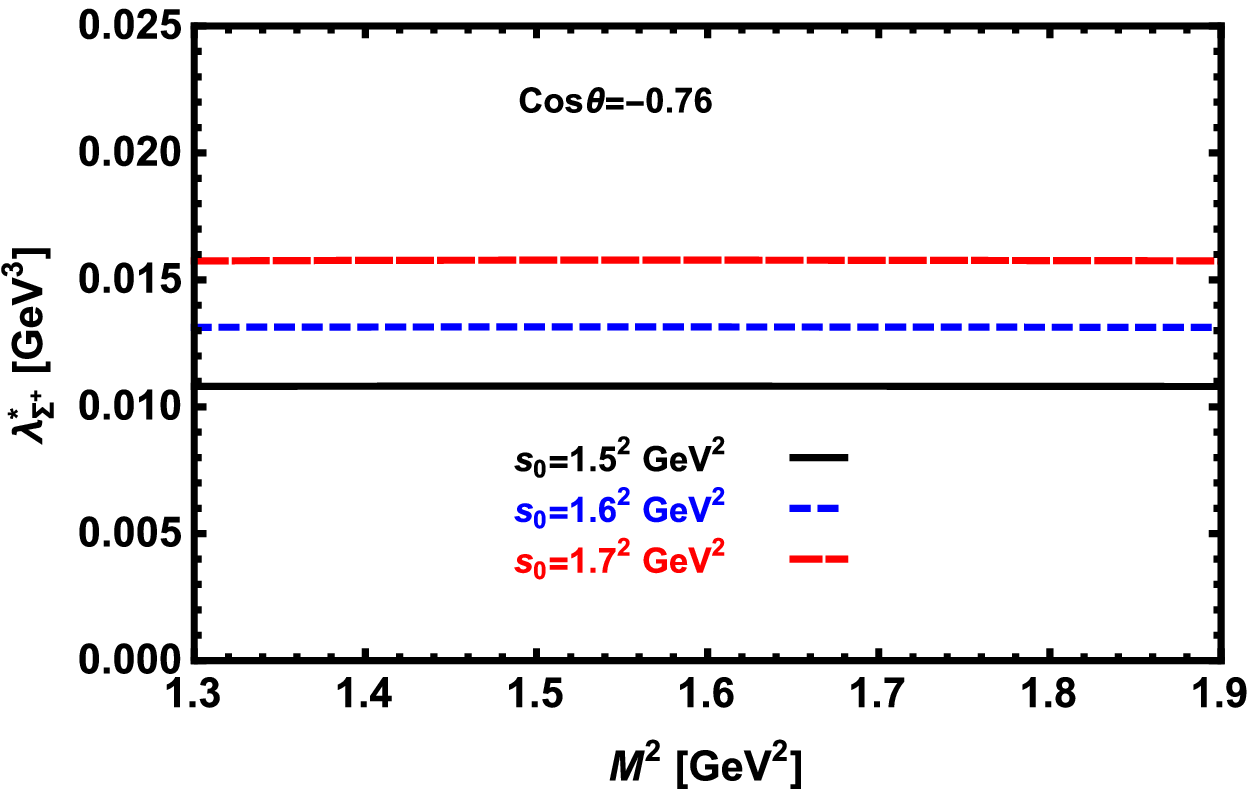,width=0.45\linewidth,clip=} &
\epsfig{file=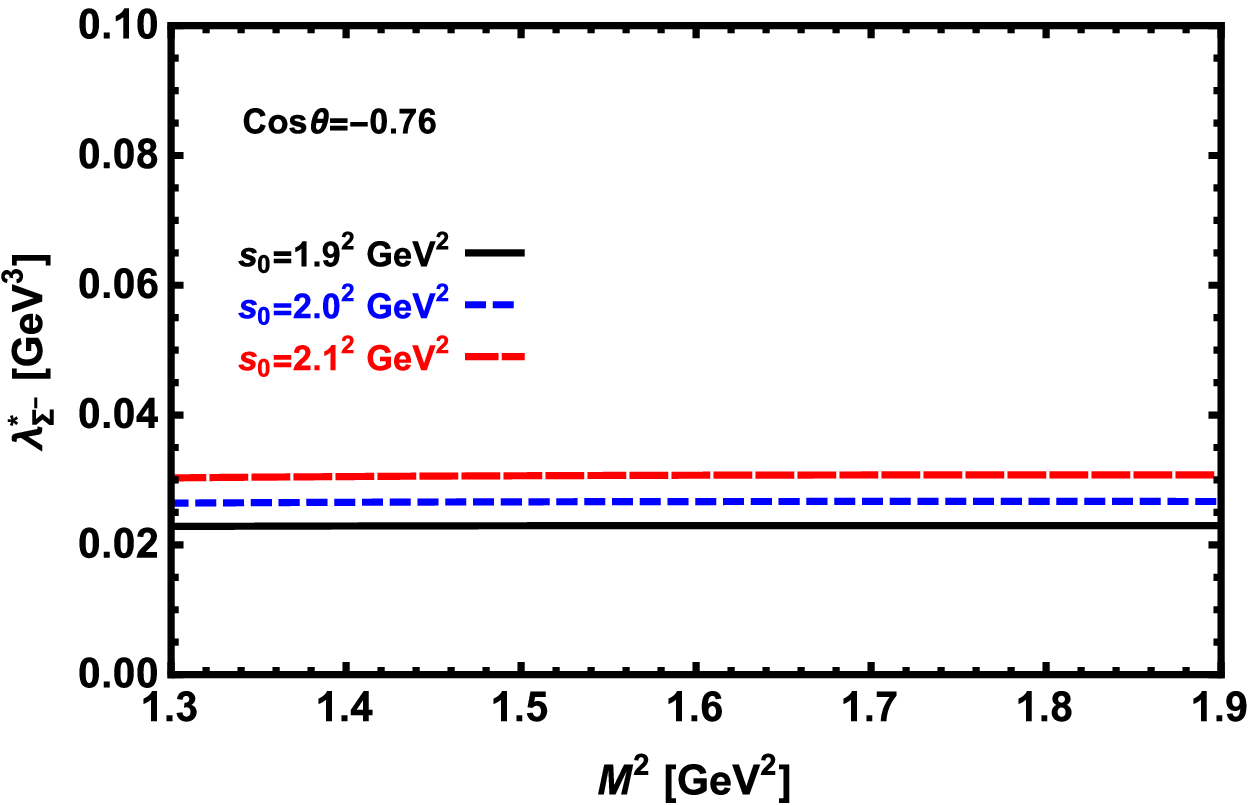,width=0.45\linewidth,clip=}
\end{tabular}
\caption{The residue of the positive parity $\Sigma$ hyperon versus  $M^2$  in nuclear matter (left panel). The same for negative parity  $\Sigma$ hyperon (right panel).}
\end{figure}

%%%%%%%%%%%%% Mass Nuc mat %%%%%%%%%%%%%

\begin{figure}[h!]
\label{fig1}
\centering
\begin{tabular}{ccc}
\epsfig{file=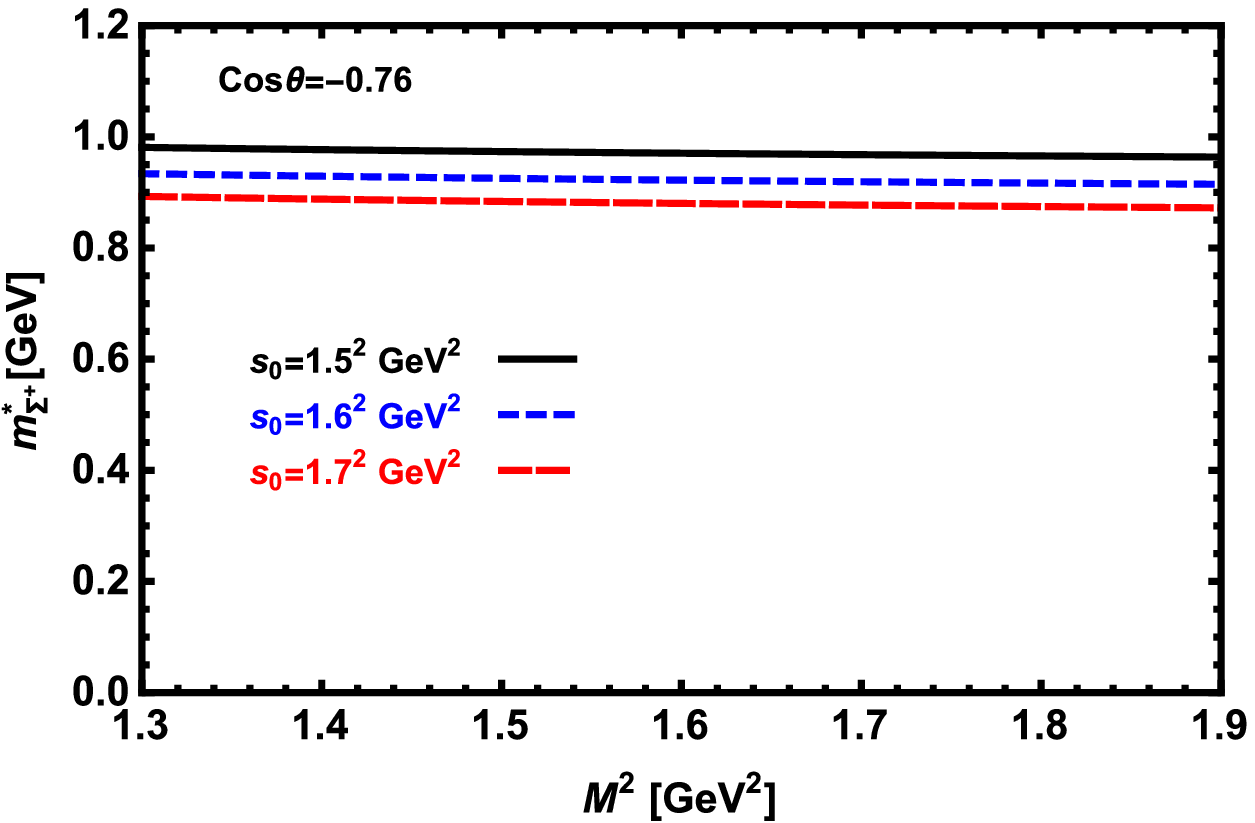,width=0.45\linewidth,clip=} &
\epsfig{file=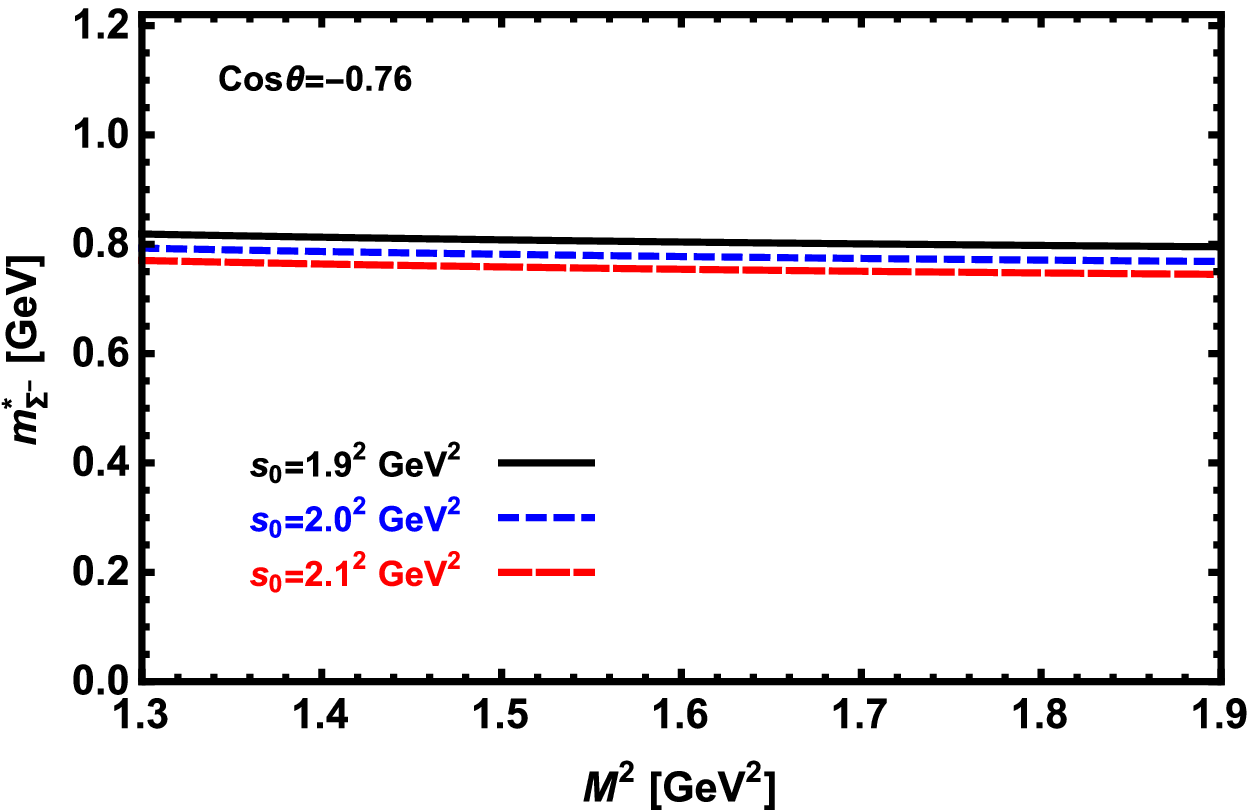,width=0.45\linewidth,clip=} 
\end{tabular}
\caption{The mass of the positive parity $\Sigma$ hyperon versus  $M^2$  in nuclear matter (left panel). The same for negative parity  $\Sigma$ hyperon (right panel).}
\end{figure}

%%%%%%%%%%%%% Self Energy  %%%%%%%%%%%%%

\begin{figure}[h!]
\label{fig1}
\centering
\begin{tabular}{ccc}
\epsfig{file=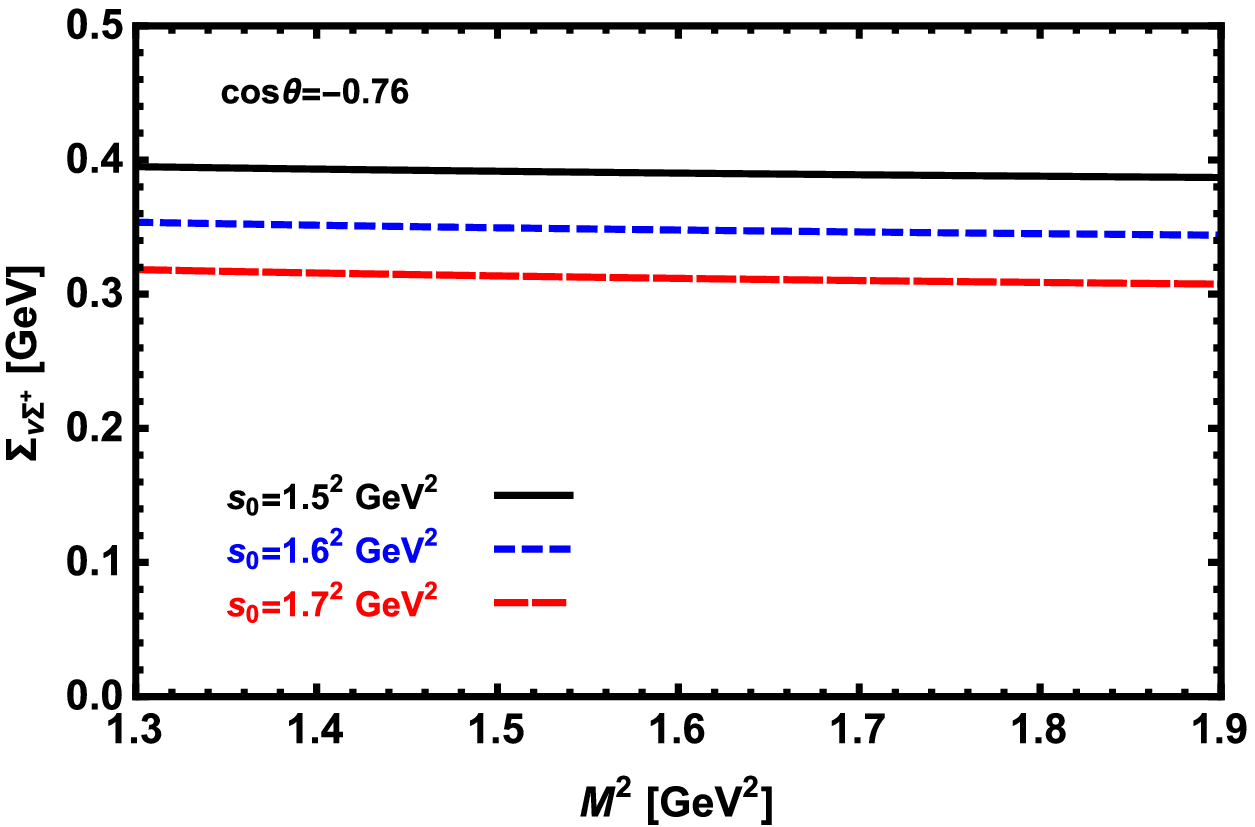,width=0.45\linewidth,clip=} &
\epsfig{file=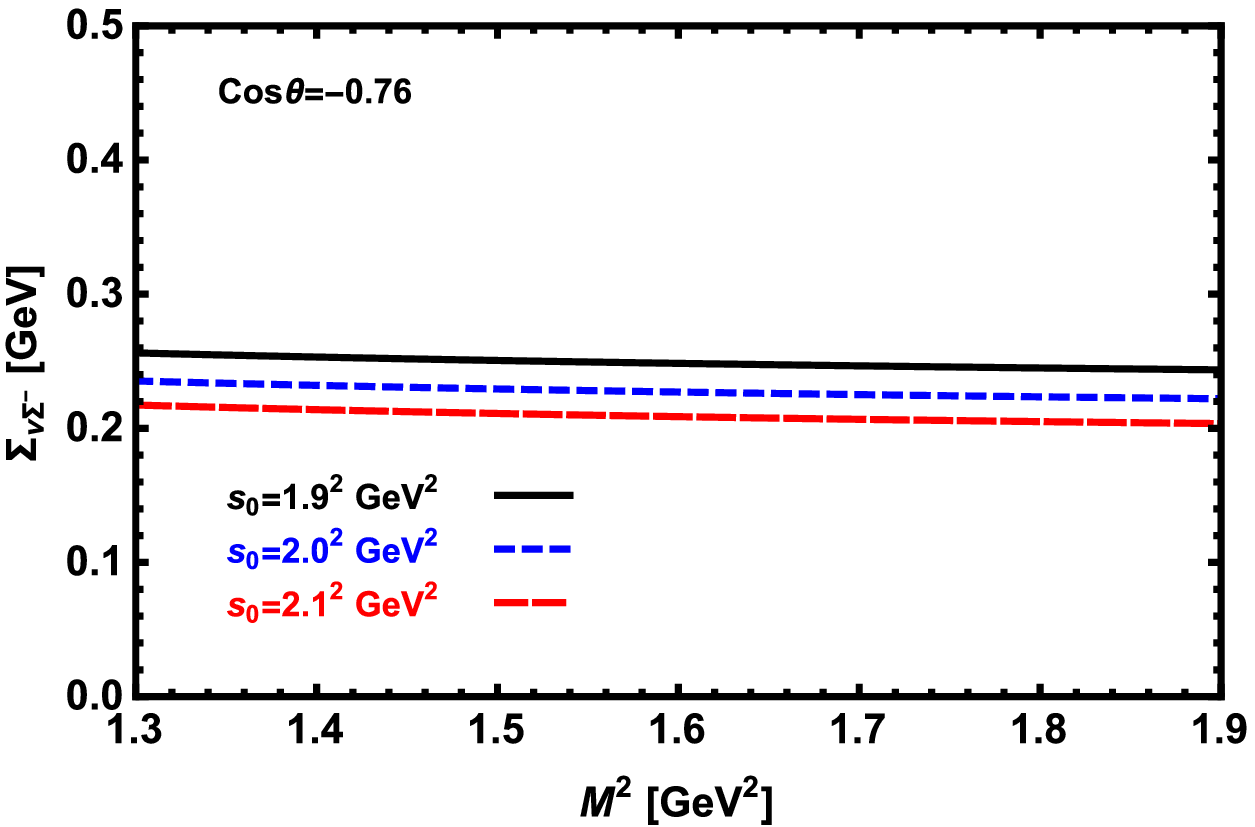,width=0.45\linewidth,clip=} 
\end{tabular}
\caption{The self-energy of the positive parity $\Sigma$ hyperon versus  $M^2$  in nuclear matter (left panel). The same for negative parity  $\Sigma$ hyperon (right panel).}
\end{figure}

%%%%%%%%%%%%%%%%%%%%%%%%%%%%%%%%%%

\subsection{$\Lambda$ hyperon}

For the $\Lambda$ hyperon, the interval of the Borel mass squared parameter is the same with $\Sigma$ hyperon, i.e.  $1.2 ~ GeV^{2}\leqslant M^2 \leqslant 1.8 ~  GeV^2$.  
For this channel we find  $cos\theta=-0.40$ demanding that the ratio residue/continuum is largest.
By considering these values, we depict the dependence
of the residues, masses and vector self-energies of the positive and negative parity $\Lambda$ particle on
the Borel mass parameter in figures 4-6. As expectedly, the behavior is similar to  the $\Sigma$ particle and we see good stabilities with
respect to $M^2$ and the continuum threshold $s_0$ at their working regions. The numerical results gained from the figures 4-6 for the average values of the residues, masses and vector
self-energies of the positive and negative $\Lambda$ hyperon are presented in table 4.
The ratios of the residues and masses in nuclear matter to
those of the vacuum together with the ratios of  the self-energies to the vacuum masses are listed in
table 5. In this table, we also depict the predictions existing in the literature for the positive
parity baryon. Similar to $\Sigma$ channel results, this table shows that there are considerable shifts in the 
parameters under consideration in nuclear matter compared to their vacuum values for
both parities in $\Lambda$ channel. The shift on the residues are positive while those are negative
for the masses. The obtained result for the ratio of the masses for the positive parity baryon in the present work is close to
that of \cite{Cohen} within the errors. 
In the comparison of the  positive and negative parity parameters, the results show
that the shifts in the residue and mass of the negative parity are considerably more than those of the positive
parity $\Lambda$. When we compare the ratio $\Sigma_{\nu \Lambda^{+}}/m_{\Lambda^{+}}$ to the ratio $\Sigma_{\nu \Lambda^{-}}/m_{\Lambda^{-}}$, we find roughly $1.36$ times differences. And finally, using the relation between the modified mass $m_{\Lambda}^{*}$ and the vacuum mass $m_{\Lambda}$, we can easily find the scalar self energy of the positive and the
negative parity $\Lambda$ to be $-0.102$ GeV and $-0.351$ GeV, respectively. Our results for the remaining
parameters can be verified by other phenomenological approaches as well as the future
experiments.

\begin{table}[ht!]
\centering
%\rowcolors{1}{lightgray}{white}
\begin{tabular}{|c|c|c|c|c|c|}
\hline \hline
 $\lambda^{*}_{\Lambda^{+}}$ [GeV$^3$] & $\lambda^{*}_{\Lambda^{-}}$ [GeV$^3$] & $m^{*}_{\Lambda^{+}}$ [GeV] & $m^{*}_{\Lambda^{-}}$ [GeV] & $\Sigma^{+}_{\nu \Lambda}$  [GeV]  &$\Sigma^{-}_{\nu \Lambda}$ [GeV] \\
 \hline
 $0.024\pm 0.007$ & $0.038 \pm 0.008$ & $ 1.022\pm 0.318 $ & $ 1.053 \pm 0.158$ & $ 0.193 \pm 0.015 $ & $0.151 \pm 0.052$ \\ 
 \hline \hline
\end{tabular}
\caption{The numerical values of residues, masses and self energies of the $\Lambda$ hyperon with positive and negative parities. }
\end{table}

\begin{table}[ht!]
\centering
%\rowcolors{1}{lightgray}{white}
\begin{tabular}{|c|c|c|c|c|c|c|}
\hline \hline
 & $\lambda^{*}_{\Lambda^{+}}/\lambda_{\Lambda^{+}}$ & $\lambda^{*}_{\Lambda^{-}}/\lambda_{\Lambda^{-}}$ & $m^{*}_{\Lambda^{+}}/m_{\Lambda^{+}}$ 
 & $m^{*}_{\Lambda^{-}}/m_{\Lambda^{-}}$ &  $\Sigma_{\nu \Lambda^{+}}/m_{\Lambda^{+}}$ &  $\Sigma_{\nu \Lambda^{-}}/m_{\Lambda^{-}}$\\
 \hline
Present work & $1.87\pm 0.36$ & $2.93\pm 0.53 $&$  0.91 \pm 0.17$&$ 0.75\pm 0.13 $&$ 0.17\pm 0.04 $&$ 0.11\pm 0.02$\\ 
 \hline
\cite{Cohen} &  &  &  0.85-0.94 & & & \\
 \hline \hline
\end{tabular}
\caption{ The ratio of parameters for  the $\Lambda$ hyperon.}
\end{table}

%%%%%%%%%%%%% Residuee Nuc mat  %%%%%%%%%%%%%

%\begin{figure}[h!]
%\label{fig1}
%\centering
%\begin{tabular}{ccc}
%\epsfig{file=LamResiduePozitiveParityNMx.eps,width=0.45\linewidth,clip=} &
%\epsfig{file=LamResidueNegativeParityNMx.eps,width=0.45\linewidth,clip=} 
%\end{tabular}
%\caption{The residue of the positive parity $\Lambda$ hyperon versus  $cos\theta$  in nuclear matter (left panel). The same for negative parity  $\Lambda$ hyperon (right panel).}
%\end{figure}

\begin{figure}[h!]
\label{fig1}
\centering
\begin{tabular}{ccc}
\epsfig{file=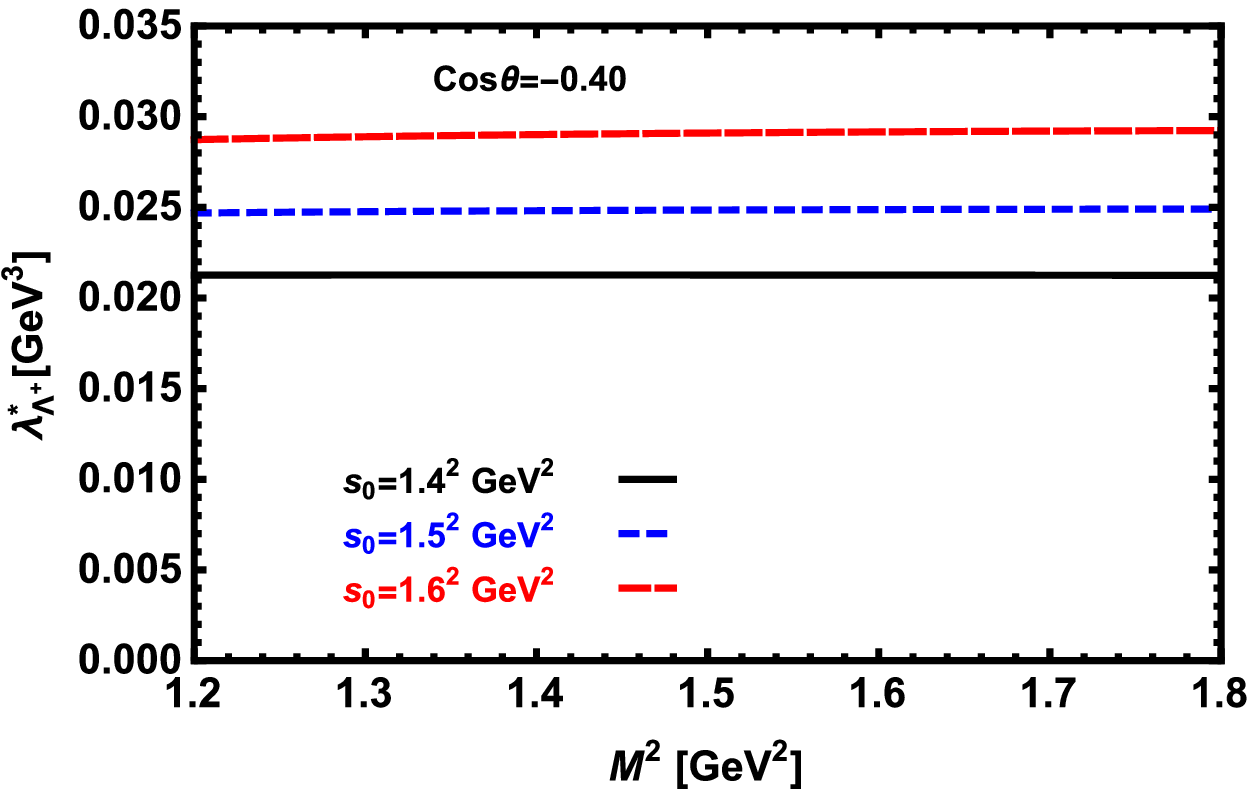,width=0.45\linewidth,clip=} &
\epsfig{file=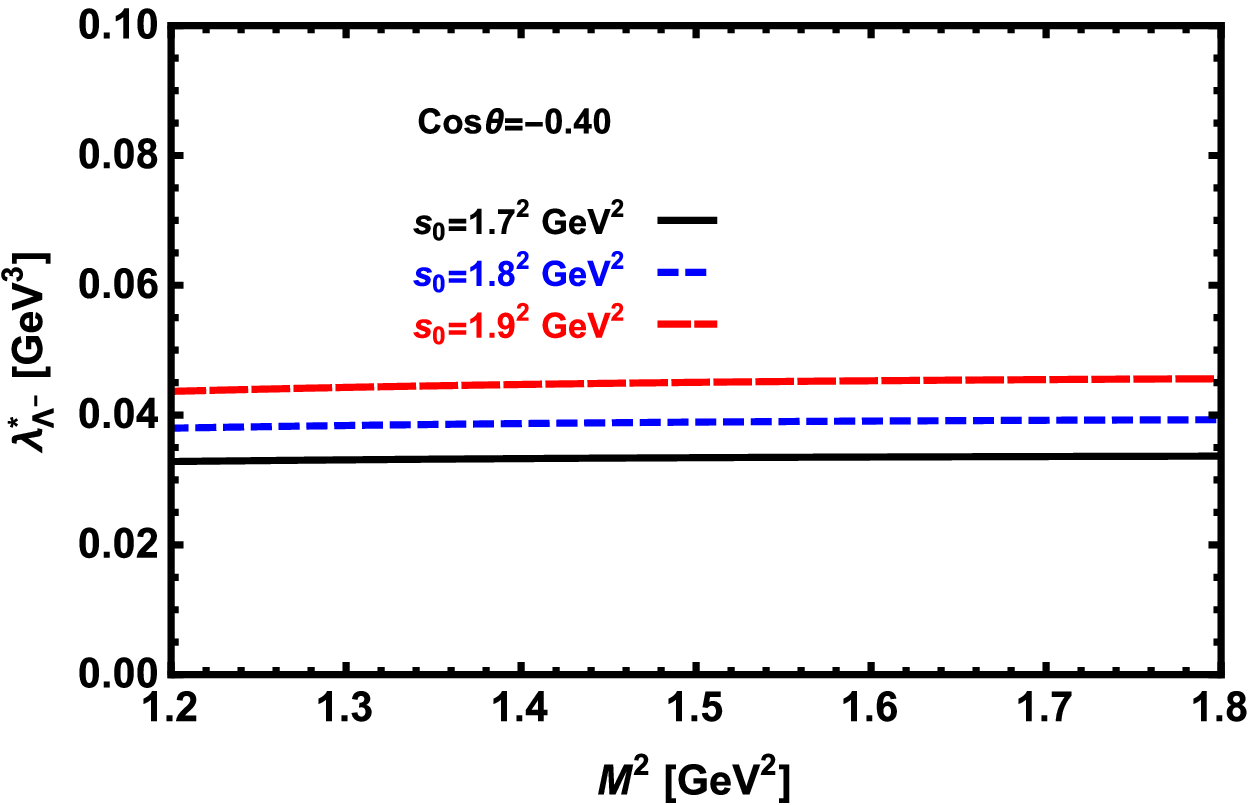,width=0.45\linewidth,clip=} 
\end{tabular}
\caption{The residue of the positive parity $\Lambda$ hyperon versus  $M^2$  in nuclear matter (left panel). The same for negative parity  $\Lambda$ hyperon (right panel).}
\end{figure}

%%%%%%%%%%%%% Mass Nuc mat  %%%%%%%%%%%%%

\begin{figure}[h!]
\label{fig1}
\centering
\begin{tabular}{ccc}
\epsfig{file=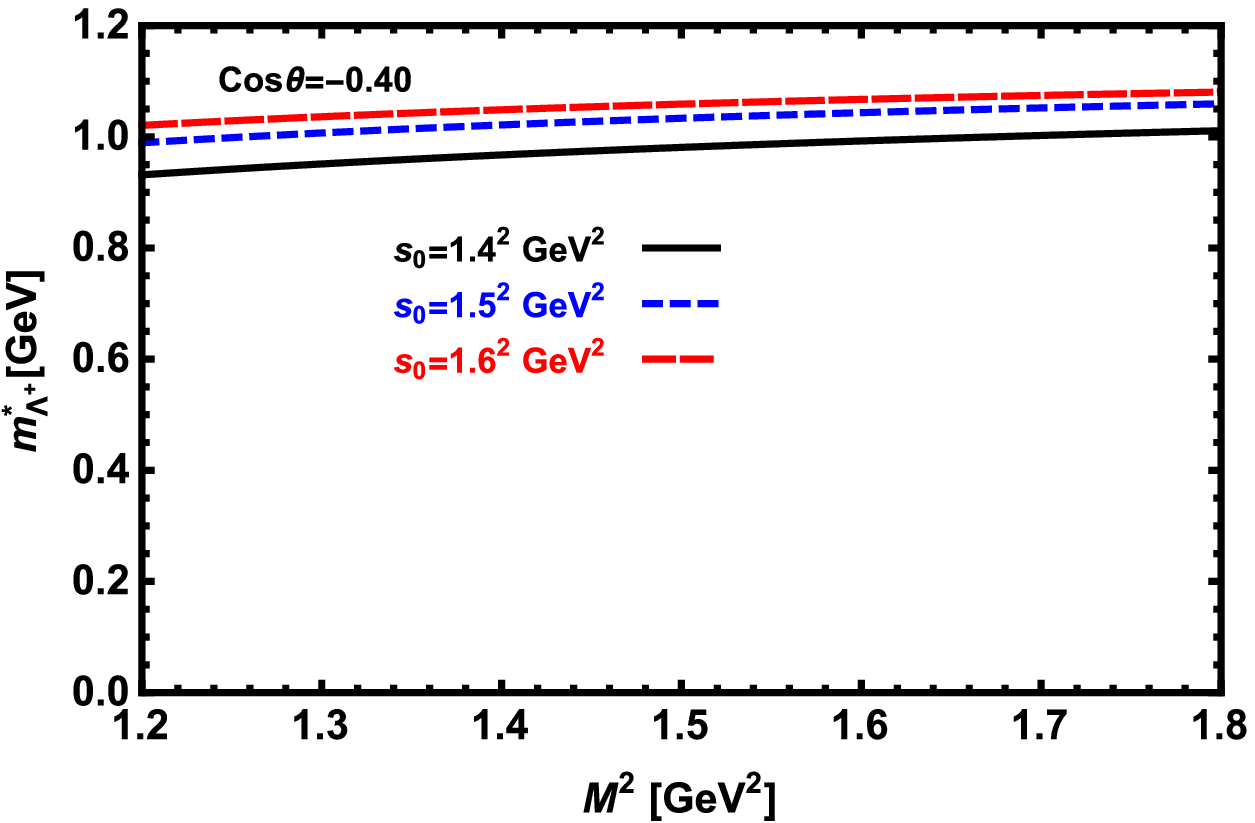,width=0.45\linewidth,clip=} &
\epsfig{file=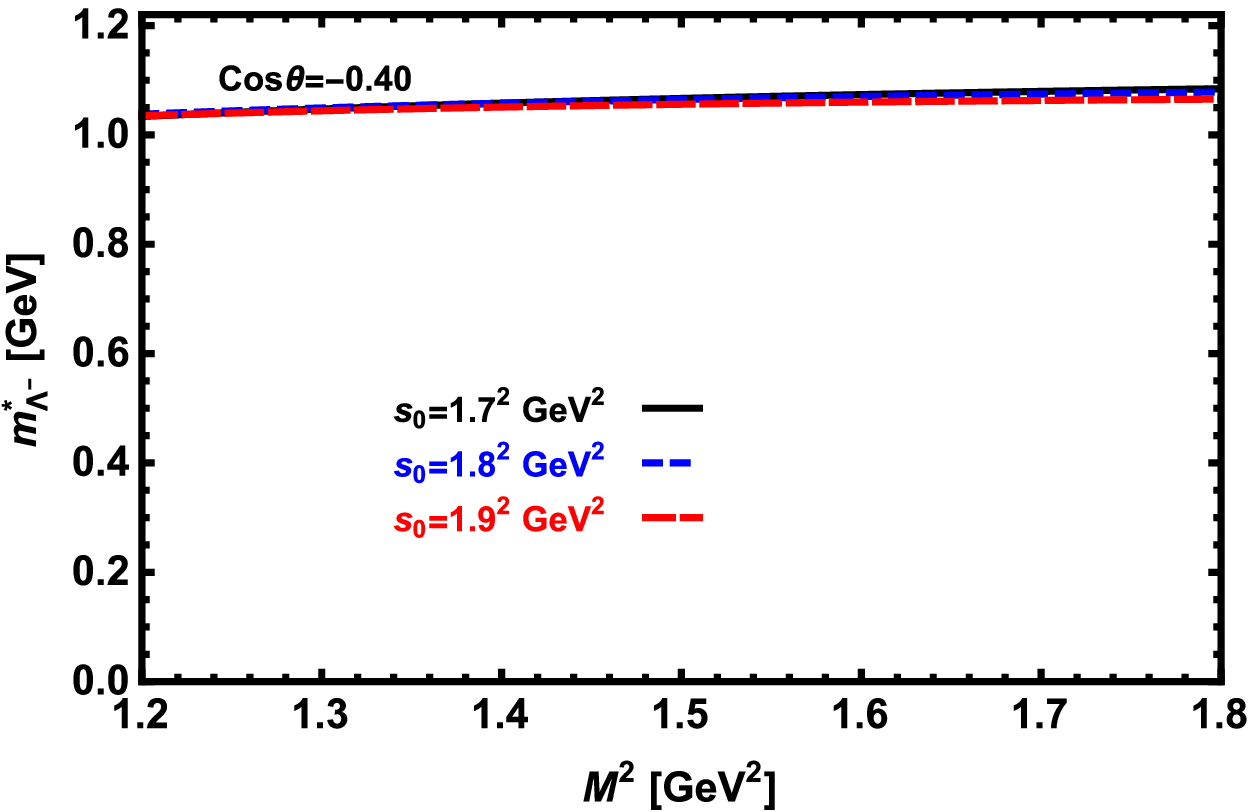,width=0.45\linewidth,clip=} &
\end{tabular}
\caption{The mass of the positive parity $\Lambda$ hyperon versus  $M^2$  in nuclear matter (left panel). The same for negative parity  $\Lambda$ hyperon (right panel).}
\end{figure}

%%%%%%%%%%%%% Self Energy  %%%%%%%%%%%%%

\begin{figure}[h!]
\label{fig1}
\centering
\begin{tabular}{ccc}
\epsfig{file=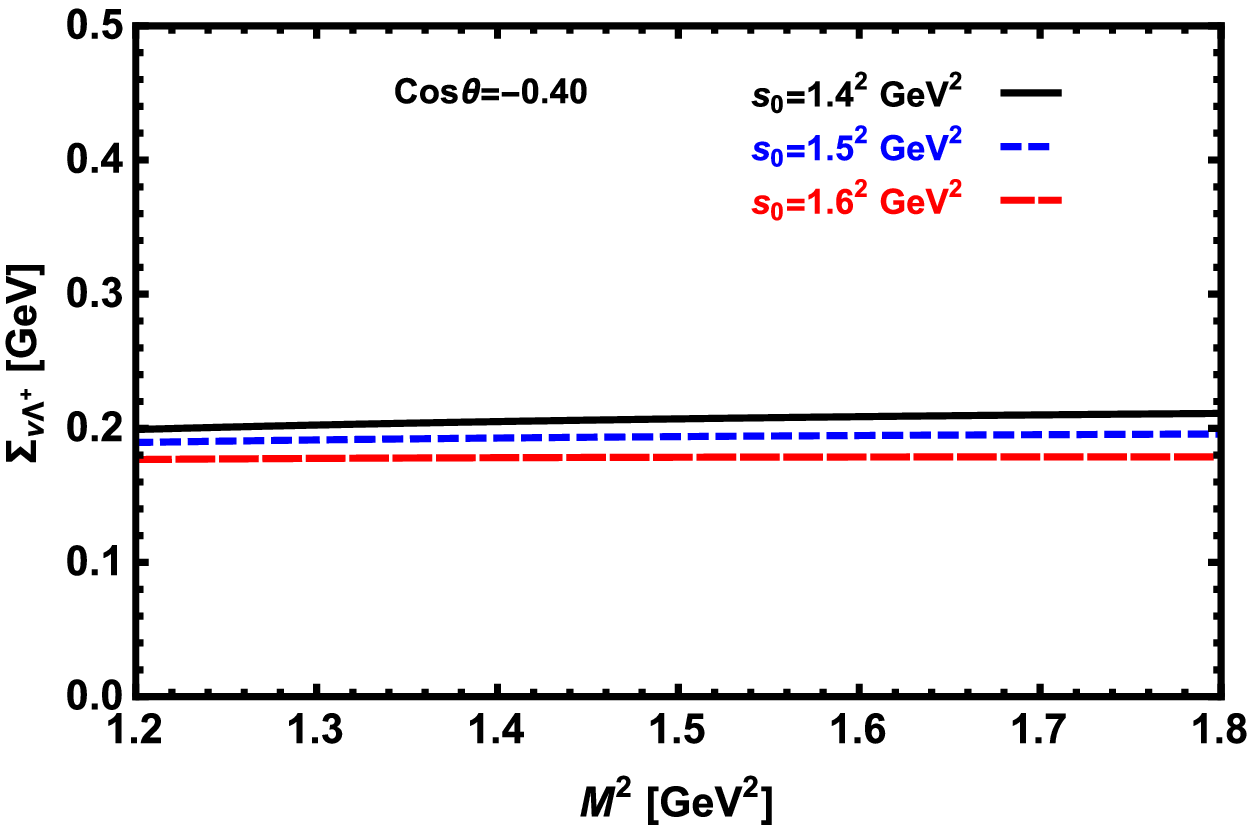,width=0.45\linewidth,clip=} &
\epsfig{file=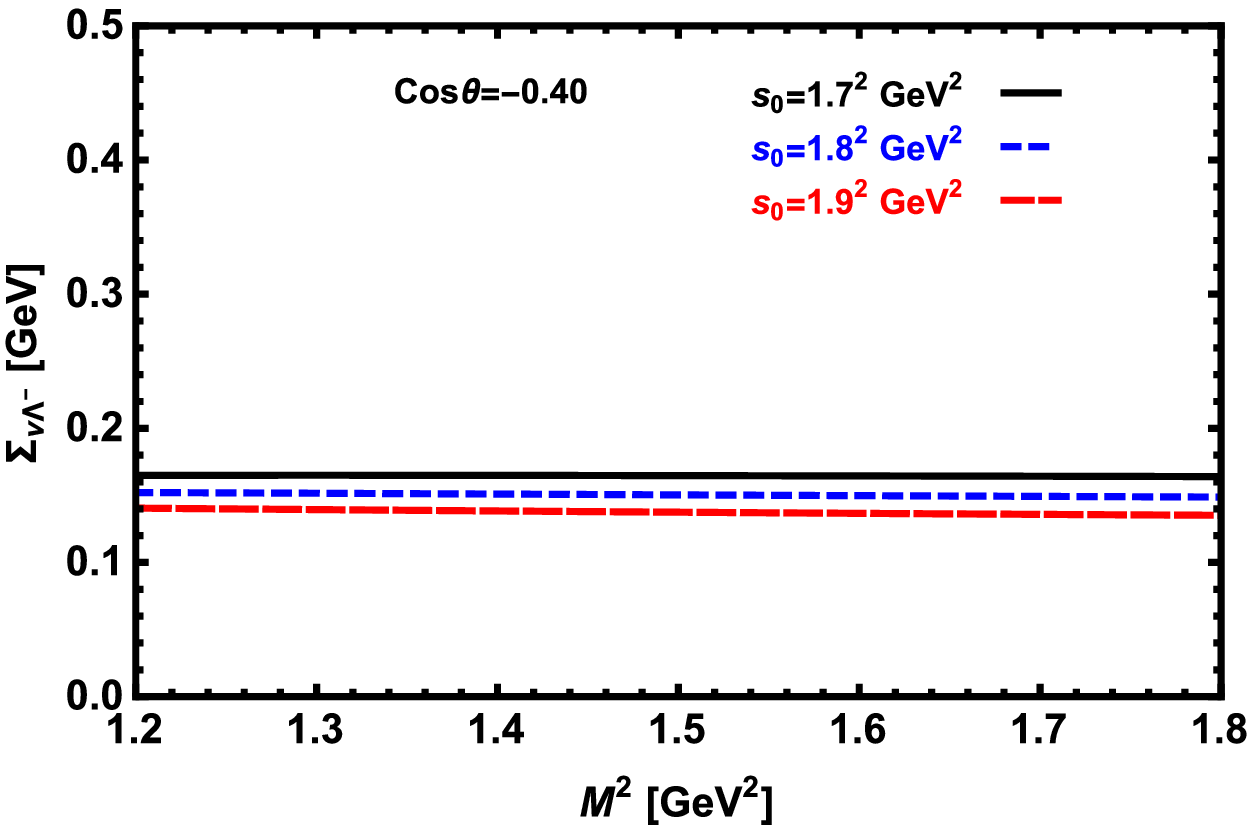,width=0.45\linewidth,clip=} 
\end{tabular}
\caption{The self-energy of the positive parity $\Lambda$ hyperon versus  $M^2$  in nuclear matter (left panel). The same for negative parity  $\Lambda$ hyperon (right panel).}
\end{figure}

%%%%%%%%%%%%%

\subsection{$\Xi$ hyperon}
In the $\Xi$ hyperon channel,  the interval  $1.5 ~ GeV^{2}\leqslant M^2 \leqslant 2.2 ~  GeV^2$ is chosen as the optimum region for the Borel mass squared parameter. 
We also  find $cos\theta=-0.43$ for this channel in order to get highest residue/continuum ratio. We show the dependence
of the residues, masses and vector self-energies of the positive and negative parity $\Xi$ particle on
the Borel mass parameter in figures 7-9.  The numerical results acquired  from these figures  for 
the average values of the residues, masses and vector
self-energies of the positive and negative parity $\Xi$ hyperon are presented in table 6.
The ratio of residues and masses in nuclear matter to
those of the vacuum together with the ratio of self-energies to the vacuum masses are also demonstrated in
table 7.  With a quick look on this table, we obtained that there are also considerable shifts in the 
parameters under consideration in nuclear matter compared to their vacuum values for
both parities in $\Xi$ channel. The shift on the residues are again positive while those are negative
for the masses. When we compare the results of parameters of the positive and negative parity cases, we see that the shifts in the residue and mass of the negative parity are considerably more than those of the positive
parity $\Xi$. With the help of the relation between $m_\Xi^*$ and $m_\Xi $ , it can be easily obtained that the scalar self energies of the positive and the
negative parity $\Xi$ are $-0.116$ GeV and $-0.504$ GeV, respectively. Our results on this channel also  can be  be tested via other approaches and experiments.

\begin{table}[ht!]
\centering
%\rowcolors{1}{lightgray}{white}
\begin{tabular}{|c|c|c|c|c|c|}
\hline \hline
 $\lambda^{*}_{\Xi^{+}}$ [GeV$^3$] & $\lambda^{*}_{\Xi^{-}}$ [GeV$^3$] & $m^{*}_{\Xi^{+}}$ [GeV] & $m^{*}_{\Xi^{-}}$ [GeV] & $\Sigma_{\nu \Xi^{+}}$ 
 [GeV] &$\Sigma_{\nu \Xi^{-}}$ [GeV] \\
 \hline
  $0.034\pm 0.009$ & $0.065\pm 0.001$ & $1.179 \pm 0.351$ & $ 1.124\pm 0.330$ & $ 0.086 \pm 0.026$ & $0.044 \pm 0.014$ \\ 
 \hline \hline
\end{tabular}
\caption{The numerical values of the residues, masses and self energies of $\Xi$ hyperon with positive and negative parities.}
\end{table}

\begin{table}[ht!]
\centering
%\rowcolors{1}{lightgray}{white}
\begin{tabular}{|c|c|c|c|c|c|c|}
\hline \hline
 & $\lambda^{*}_{\Xi^{+}}/\lambda_{\Xi^{+}}$ & $\lambda^{*}_{\Xi^{-}}/\lambda_{\Xi^{-}}$ & $m^{*}_{\Xi^{+}}/m_{\Xi^{+}}$ & $m^{*}_{\Xi^{-}}/m_{\Xi^{-}}$
 &  $\Sigma_{\nu \Xi^{+}}/m_{\Xi^{+}}$ &  $\Sigma_{\nu \Xi^{-}}/m_{\Xi^{-}}$\\
 \hline
Present work & $2.05 \pm 0.37$ & $2.83 \pm0.49$&$ 0.91 \pm0.18$  & $0.69 \pm0.12$ & $0.07  \pm0.01$&$0.03 \pm0.00$\\ 
 \hline \hline
\end{tabular}
\caption{ The ratio of parameters for the $\Xi$ hyperon.}
\end{table}

%%%%%%%%%%%%% Residuee Nuc mat  %%%%%%%%%%%%%

%\begin{figure}[h!]
%\label{fig1}
%\centering
%\begin{tabular}{ccc}
%\epsfig{file=XiResiduePozitiveParityNMx.eps,width=0.45\linewidth,clip=} &
%\epsfig{file=XiResidueNegativeParityNMx.eps,width=0.45\linewidth,clip=} &
%\end{tabular}
%\caption{The residue of the positive parity $\Xi$ hyperon versus  $cos\theta$  in nuclear matter (left panel). The same for negative parity  $\Xi$ hyperon (right panel).}
%\end{figure}

\begin{figure}[h!]
\label{fig1}
\centering
\begin{tabular}{ccc}
\epsfig{file=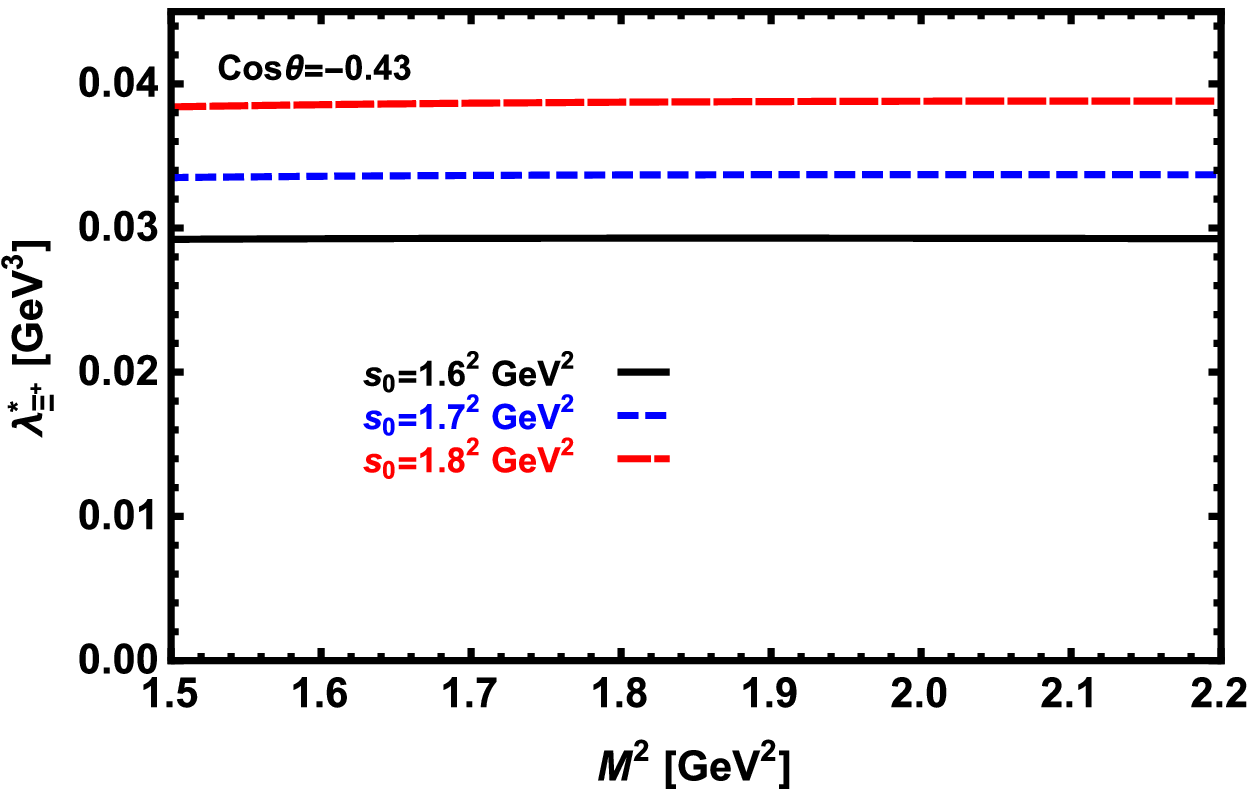,width=0.45\linewidth,clip=} &
\epsfig{file=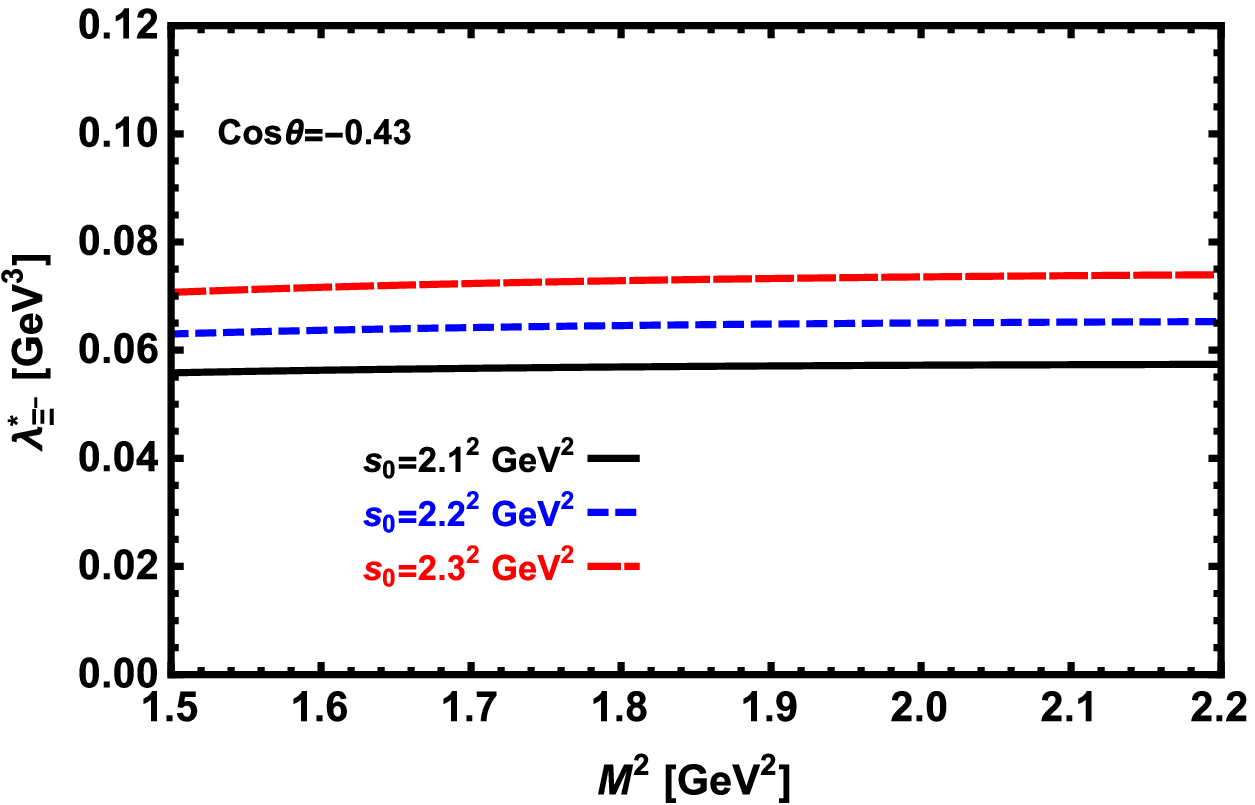,width=0.45\linewidth,clip=} 
\end{tabular}
\caption{The residue of the positive parity $\Xi$ hyperon versus  $M^2$  in nuclear matter (left panel). The same for negative parity  $\Xi$ hyperon (right panel).}
\end{figure}

%%%%%%%%%%%%% Mass Nuc mat  %%%%%%%%%%%%%

\begin{figure}[h!]
\label{fig1}
\centering
\begin{tabular}{ccc}
\epsfig{file=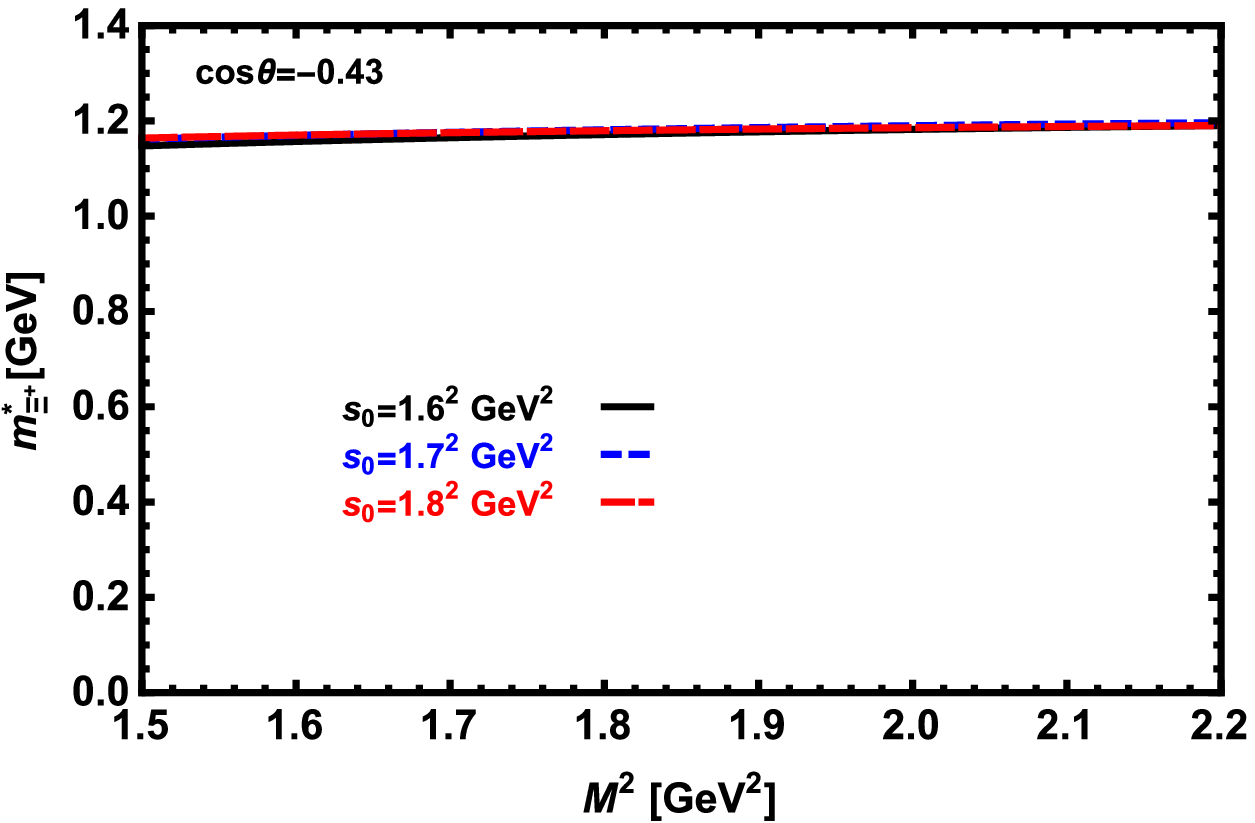,width=0.45\linewidth,clip=} &
\epsfig{file=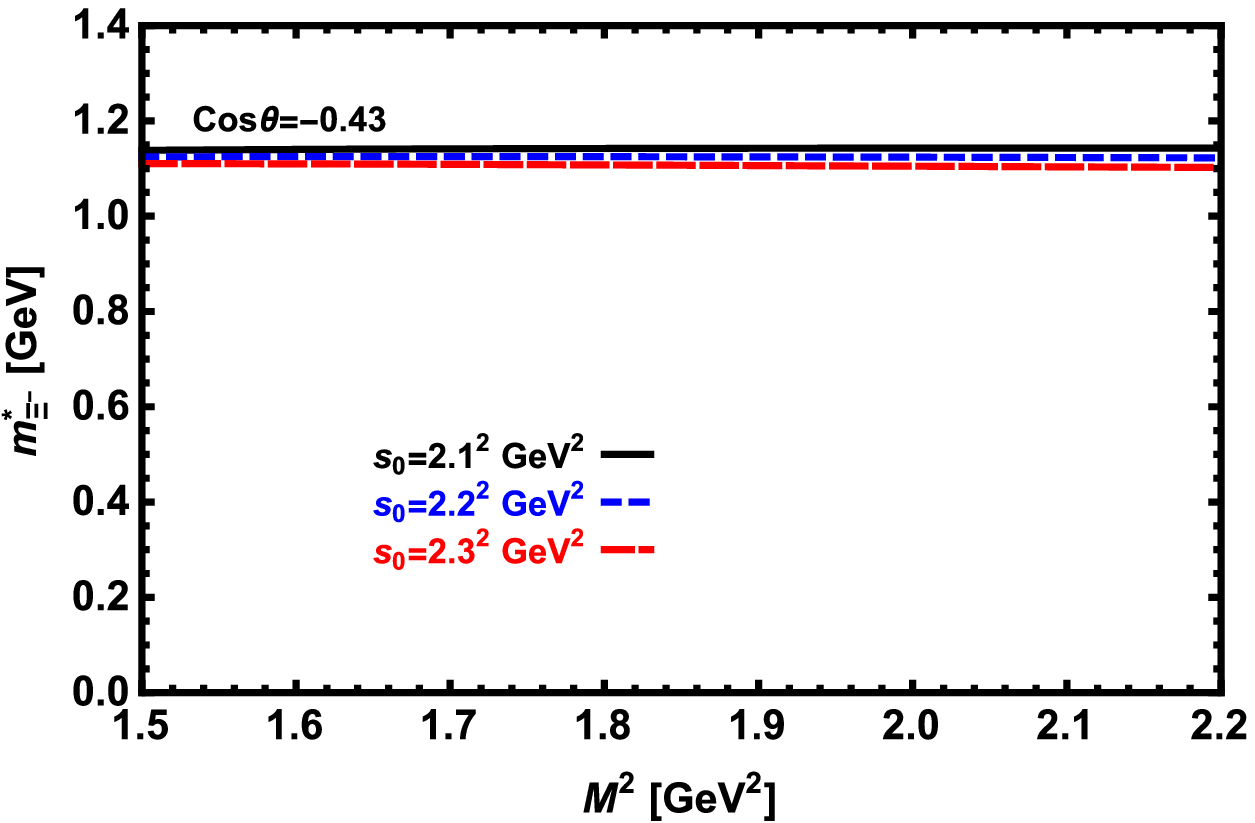,width=0.45\linewidth,clip=} 
\end{tabular}
\caption{The mass of the positive parity $\Xi$ hyperon versus  $M^2$  in nuclear matter (left panel). The same for negative parity  $\Xi$ hyperon (right panel).}
\end{figure}

%%%%%%%%%%%%% Self Energy  %%%%%%%%%%%%%

\begin{figure}[h!]
\label{fig1}
\centering
\begin{tabular}{ccc}
\epsfig{file=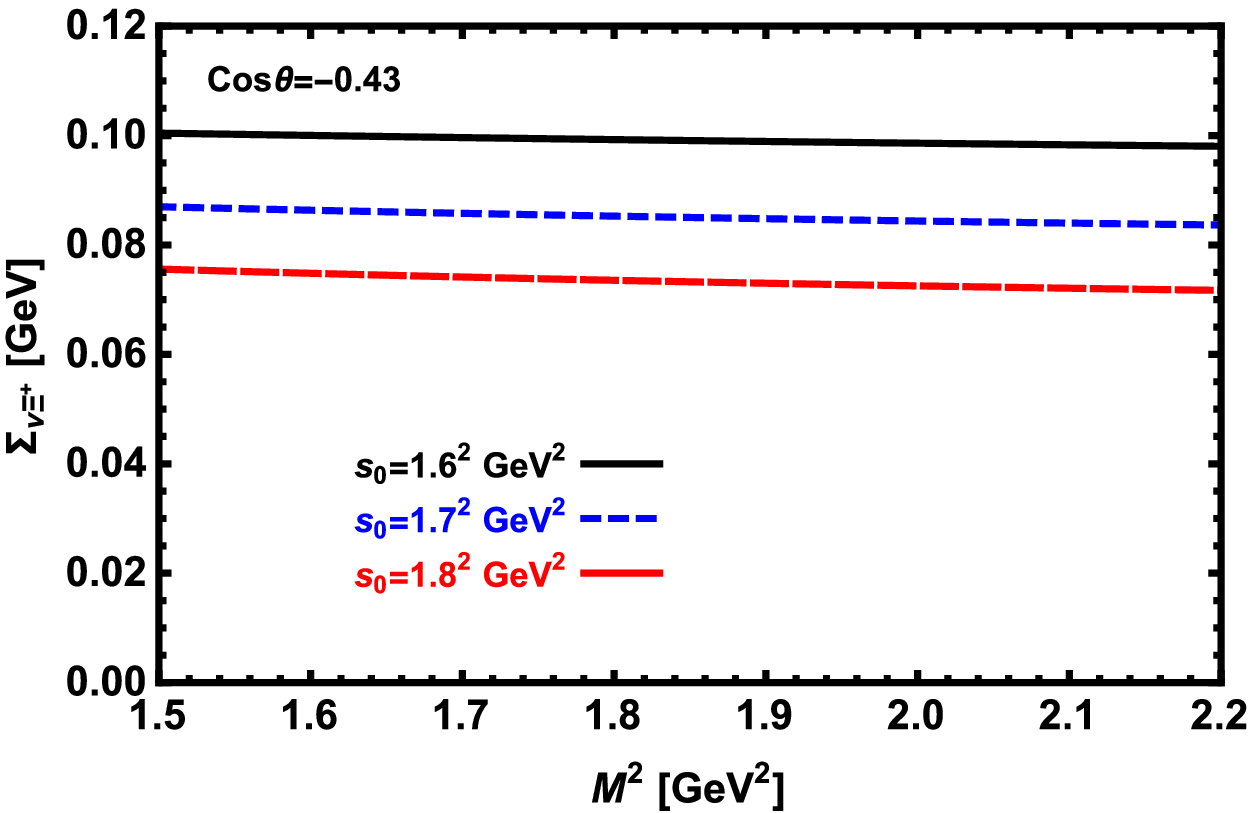,width=0.45\linewidth,clip=} &
\epsfig{file=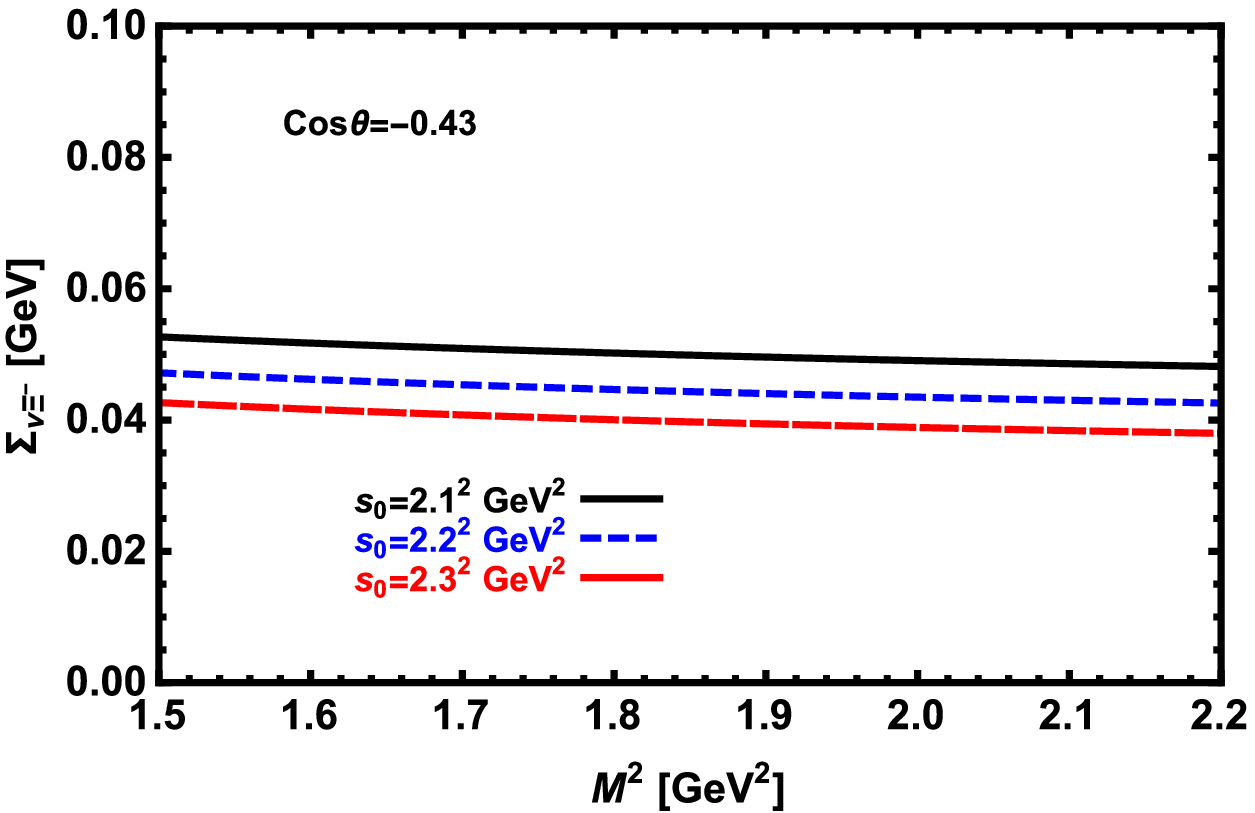,width=0.45\linewidth,clip=} 
\end{tabular}
\caption{The self-energy of the positive parity $\Xi$ hyperon versus  $M^2$  in nuclear matter (left panel). The same for negative parity  $\Xi$ hyperon (right panel).}
\end{figure}

%%%%%%%%%%%%%

\section{Concluding remarks} 
We investigated the effects of nuclear medium on the residues, masses and self-energies of the positive and negative parity $\Sigma, \Lambda$ and $\Xi$ hyperons 
in the framework of the QCD sum rule method. We used the general interpolating currents of these baryons with an arbitrary mixing parameter to calculate the shifts
 in the masses and residues of these hyperons for both the positive and negative parity states in nuclear medium compared to their vacuum values.
 We also  obtained the values of the vector and scalar self-energies for these baryons considering  the reliable working regions of the auxiliary parameters entering the calculations. 
We observed that the shifts on the residues in nuclear matter are over all positive  for both the positive and negative parity  hyperons, except for the positive parity  $\Sigma$ hyperon
 that it is negative. The shifts on the masses of 
these baryons are obtained to be  negative. It is found that the shifts on the residues and the masses
 corresponding to the negative parity particles are considerably large compared to those of
the positive parity states. The maximum shift belong to the value of the residue of the negative parity $\Lambda$  baryon.  In the case of vector self-energy, the energy gained by the 
positive parity baryons are large in comparison with those of negative parity vector self-energy. The maximum value of the vector self-energy corresponds to the positive parity $\Sigma$ hyperon. 
In the case of absolute value of the scalar self-energy, we found that the maximum and minimum values belong to the negative parity $\Sigma$ and the positive parity $\Lambda$ hyperons, respectively.  
Our results can be checked via different phenomenological approaches as well as the future experiments. The obtained results can also be used in analyses of the heavy ion collision experiments.  

\section{Acknowledgement}
This work has been supported in part by the Scientific and Technological Research Council of Turkey (TUBITAK) under the grant no: 114F018.

                                 %%%%%%%%%%%%%%%%%%%%%%%%%%%%%%%%%%%%%%%
       %%%%%%%%%%%%%%%%%%%%%%%%%%%%%%%%%%%%%%%             %%%%%%%%%%%%%%%%%%%%%%%%%%%%%%%%%%%%%%%%%%
                                 %%%%%%%%%%%%%%%%%%%%%%%%%%%%%%%%%%%%%%%

\begin{thebibliography}{99}

\bibitem{Drukarev88} E. G. Drukarev and E. M. Levin, Pis'ma Zh. Eksp. Teor. Fiz. 48, 307 (1988).

\bibitem{Drukarev90}E. G. Drukarev and E. M. Levin, Nucl. Phys. A 511, 679, (1990); 516, 715(E) (1990).

\bibitem{Hatsuda91}T. Hatsuda, H. Hogaasen, M. Prakash, Phys. Rev. Lett. 66, 2851 (1991).

\bibitem{Adami91} C. Adami, G. E. Brown, Z. Phys. A 340, 93 (1991).
 
\bibitem{kanur2014} K. Azizi, N. Er,  Eur. Phys. J. C 74, 2904 (2014).

\bibitem{Jin94} Xuemin Jin and , R. J. Furnstahl, Phys. Rev. C 49, 1190 (1994).
 
\bibitem{Jin95} Xuemin Jin, Marina Nielsen, Phys. Rev. C51, 347 (1995).

\bibitem{Savage96} Martin J. Savage, Mark B. Wise, Phys.Rev. D53, 349-354  (1996).

\bibitem{Miyatsu2009} T. Miyatsu, K. Saito, Prog. Theor. Phys. 122, 1035-1044  (2009).

\bibitem{Beane2012} S. R. Beane, E. Chang, S. D. Cohen, W. Detmold, H.-W. Lin, T. C. Luu, K. Orginos, A. Parreno, M. J. Savage, A. Walker-Loud, Phys. Rev. Lett. 109, 172001 (2012).

\bibitem{Chung} V. Chung, H. G. Dosch, M. Kremer, D. Scholl, Nucl. Phys. B 197, 55 (1982).

\bibitem{Dosch} H. G. Dosch, M. Jamin and S. Narison, Phys. Lett. B 220, 251 (1989).

\bibitem{Thomas} R. Thomas, T. Hilger, B. Kampfer, Nucl. Phys. A 795, 19 (2007).

\bibitem{Drukarev2013} E. G. Drukarev, M. G. Ryskin, V. A. Sadovnikova, arXiv:1312.1449[hep-ph].

\bibitem{Leinweber} D. B. Leinweber, Phys. Rev. D 51, 6383 (1995).

\bibitem{Stein} E. Stein, P. Gornicki, L. Mankiewicz, A. Schafer, W. Greiner, Phys. Lett. B 343, 369 (1995).

\bibitem{TDC2}  T. D. Cohen, R. J. Furnstahl, and David K. Greigel, Phys. Rev. Lett. 67, 961 (1991).

\bibitem{Cohen} T. D. Cohen, R. J. Furnstahl, D. K. Griegel and Xuemin Jin, Prog. Part. Nucl. Phys. 35, 221 (1995).
\bibitem{PDG} K.A. Olive et al. (Particle Data Group), Chin. Phys. C, 38, 090001 (2014).
\bibitem{Yoshihiko} Yoshihiko Kondo, Osamu Morimatsu, Tetsuo Nishikawa, and Yoshiko Kanada-En'yo, Phys. Rev. D 75, 034010 (2007).



\bibitem{Cohen45} T. D. Cohen, R. J. Furnstahl and D. K. Griegel, Phys. Rev. C 45, 1881 (1992).

\bibitem{XJ1}  X. Jin, T. D. Cohen, R. J. Furnstahl, and D. K. Griegel, Phys. Rev. C 47, 2882 (1993).
\bibitem{Wang} Zhi-Gang Wang, Eur. Phys. J. C72, 2099 (2012).
 \bibitem{Belyaev} V. M. Belyaev, B. L. Ioffe,  Sov. Phys. JETP 57, 716 (1983); B. L. Ioffe,Prog. Part. Nucl. Phys. 56, 232 (2006).

\bibitem{Thomas1} A. W. Thomas, P. E. Shanahan, R. D. Young, Nuovo Cim. C 035N04, 3 (2012).
\bibitem{Dinter} S. Dinter, V. Drach, K. Jansen, Int. J. Mod. Phys. Proc. Suppl. E 20, 110 (2011).


\bibitem{Nielsen} X. Jin, M. Nielsen, T. D. Cohen, R. J. Furnstahl, D. K. Griegel, Phys. Rev. C 49, 464 (1994).




%\bibitem{Savage} Martin J. Savage, Mark B. Wise, Phys.Rev. D53, 349-354  (1996).

%\bibitem{Shifman} M. A. Shifman, A. I. Vainshtein, V. I. Zakharov, Nucl. Phys. B 147, 385 (1979).

%\bibitem{Colangelo} P. Colangelo, A. Khodjamirian, ��At the Frontier of Particle Physics/Handbook of QCD“, edited by M. Shifman (World Scientific, Singapore, 2001), Vol.3, p. 1495. 

%\bibitem{Rusnak} J. J. Rusnak and R. J. Furnstahl, Ohio State University Report No. 94-220 (1994) unpublished. 


\end{thebibliography}
\end{document}